\documentclass[12pt]{article}
\usepackage{amstext,amssymb,amsfonts,amsmath,amsthm,mathtools}
\usepackage{array}
\usepackage{graphicx,subfigure}
\usepackage{textcomp} 
\usepackage{gensymb} 
\usepackage{ulem}
\graphicspath{ {./figures/} } 
\usepackage[font=footnotesize]{caption}
\usepackage{epstopdf}
\usepackage{xcolor}
\usepackage{soul}
\usepackage{url}
\usepackage{enumerate}
\usepackage[toc,page]{appendix}
\usepackage[margin=1.0in]{geometry}
\usepackage{environ}
\NewEnviron{SMALL}{\scalebox{0.90}{$\BODY$}}
\NewEnviron{SSMALL}{\scalebox{0.50}{$\BODY$}}

\setulcolor{red}
\begin{document}
\begin{center}
\Large{\textbf{Simplified Two-Dimensional Model for Global Atmospheric Dynamics}}\footnote{This is an extended version of preprint \cite{Jacques-Coper:2021lvg}}\\
\vskip 5.mm
Mart\'{\i}n Jacques-Coper$^{1,2,3,\dag}$, Valentina Ortiz-Guzm\'an${}^{4,5\ddag}$ and Jorge Zanelli${}^{4,6\sharp}$ \\
\vskip 0.5cm
\small{${}^{1}$Departamento de Geof\'{\i}sica, Universidad de Concepci{\'o}n, Casilla 160-C, Concepci{\'o}n, Chile} \\ 
\small{${}^{2}$ Center for Climate and Resilience Research (CR2), Universidad de Concepci{\'o}n, Concepci{\'o}n, Chile} \\ 
\small{${}^{3}$ Center for Oceanographic Research COPAS Coastal, Universidad de Concepci{\'o}n, Concepci{\'o}n, Chile}\\
\small{${}^4$ Centro de Estudios Cient\'{\i}ficos (CECs), Av. Arturo Prat 514, Valdivia, Chile}\\
\small{${}^5$ Climate Change Research Centre, University of New South Wales, Sydney, Australia}\\
\small{${}^6$ Universidad San Sebasti{\'a}n, Av. General Lagos 1163, Valdivia, Chile.}
\vskip 0.2cm
\texttt{\footnotesize{$^{\dag}$mjacques@dgeo.udec.cl, $^{\ddag}$vaortizguzman@gmail.com, $^{\sharp}$jorge.zanelli@uss.cl}}\\
\vskip 0.5cm
{\bf Abstract}
\end{center}
We present a simplified model of the atmosphere of a terrestrial planet as an open two-dimensional system described by an ideal gas with velocity $\vec{v}$, density $\rho$ and temperature $T$ fields. Starting with the Chern-Simons equations for a free inviscid fluid, the external effects of radiation and the exchange of matter with the strata, as well as diffusion and dissipation are included. The resulting dynamics is governed by a set of nonlinear differential equations of first order in time. This defines an initial value problem that can be integrated given the radiation balance of the planet. If the nonlinearities are neglected, the integration can be done in analytic form using standard Green function methods, with small nonlinearities incorporated as perturbative corrections in a consistent way. If the nonlinear approximation is not justified, the problem can be integrated numerically. The analytic expressions as well as the simulations of the linear regime for a continuous range of parameters in the equations are provided, which allows to explore the response of the model to changes of those parameters. In particular, it is observed that a 2.5\% reduction in the emissivity of the atmosphere can lead to an increase of 7$\degree$C of the average global temperature.\\

\section{Introduction}  

The study of Earth's climate has been a subject of interest since at least the early 19th century. However, it has raised major global concern during recent decades, mainly due to the impact of anthropogenic global warming. The 6th Assessment Report of the IPCC \cite{ipcc2021} states that, during such period, ``it is unequivocal that human influence has warmed the atmosphere, ocean and land" and that ``strong, rapid, and sustained reductions" in greenhouse gases are required ``to limit human-induced global warming to a specific level". In a geological timescale, it is relevant to analyse the possible transitions between Earth?s climate states in the past, present, and future due to natural and human forcing, considering trajectories leading to both unstable states, such as the so-called ``Hothouse Earth" \cite{steffen2018} and stable states, such as the current warm climate and snowball climate (global glaciation) \cite{lucarini2017}. Unstable climate states have received further attention \cite{lucarini2019}. In this context, simplified climate models are contributions to the understanding of the global, steady-state climate behaviour under different perturbations.

The time-honored approach to atmospheric dynamics models it as a fluid obeying Newtonian laws to describe the changes in momentum, energy and density of a parcel \cite{Lorenz1969}. Historically, different approximations have been followed to derive models for the global atmosphere \cite{White2005}. A typical and useful form of such a model are The Shallow Water Equations (SWE), used when describing a system where the horizontal length scale is much greater than the vertical length scale. The dynamical equations of non-dissipative systems can be usually derived from an action principle through the Euler-Lagrange equations, which in simple cases reduce to Newton's Laws. Even if the dynamical equations obtained by two distinct approaches turn out to be equivalent in simple cases, there may be settings  where one approach can be more useful or more appropriate than the other. One set of assumptions may reflect better the conditions or may simplify the presentation to make the description more transparent.

 Fluid equations are essentially derived from Newton's second law locally applied to a fluid element, with the addition of particular features such as viscosity, compressibility, thermodynamics and interactions with external sources. An alternative equivalent to Newtonian dynamics when the forces involved are derived from a potential is the principle of least action, in which case Newton's laws are obtained in the form of Lagrange's or Hamilton's equations. Effects such as friction or viscosity producing dissipation --typically in the form of heat--, can also be accounted for in the dynamical equations when the mechanisms that transform kinetic and potential energy into heat are clearly understood. The advantage of having a larger picture such as an action principle, is that it can cover a wider range of phenomena within the same framework. An action principle allows a better understanding of symmetries, conservation laws, integrability, quantization and phase transitions in a classical dynamical system.

In the Lagrangian approach, the evolution is described through second order equations in time of the form $\ddot{\vec{x}} = F(\vec{x},\dot{\vec{x}},t)$. The Hamiltonian approach produces equivalent equations but as a first order system, typically of the form $\dot{\vec{p}}=F(\vec{x},\vec{p},t)\,,\; \vec{p}=m\dot{\vec{x}}$. The solutions of the Lagrangian equations are all possible trajectories starting from any initial position $\vec{x}|_0$ with any velocity $\dot{\vec{x}}|_0$, which is usually a complicated tangle of trajectories in $(\vec{x},t)$ space. In the Hamiltonian approach, those trajectories in configuration space are replaced by the motion of a fluid in $(\vec{x},\vec{p})$  in phase space (Hamiltonian flow). An additional advantage of the Hamiltonian approach is the fact that the momentum $\vec{p}$ allows for changes in mass which, in the case of a real fluid, correspond to changes in density produced by compression and temperature variations.

It can be shown that if the even-dimensional phase space of a Hamiltonian system is viewed as a manifold with coordinates $z^i$, $i=1,2,\cdots 2n$, where a vector field $A_\mu(z), \; (\mu=0, 1, 2,\cdots 2n)$ takes values, then Hamilton's equations can be cast as those of a Chern-Simons (\textbf{CS}) system \cite{UsesofCS}. Chern-Simons actions are particularly apt to describe $2n$-dimensional fluids. The simplest CS model describes the evolution in time of a 2-dimensional fluid and has been shown to provide a good model for the quantum Hall effect \cite{Frohlich} electrons in the two-dimensional planes of high-temperature superconductors \cite{RandjbarDaemi:1989ds, Sedrakyan:2016udf, Wang:2020zwt}, or in the two-dimensional carbon lattice of graphene \cite{AVZ, Andrianopoli}. 

As will be seen here, the CS description naturally selects as relevant variables the momentum density, $A_i\sim \rho v_i$ instead of velocity; heat energy density, $A_0\sim \rho\,T$ instead of temperature; etc. Similarly, the CS dynamics links the vorticity to the changes in density and produces a seemingly reasonable account of the phenomena in the linearized approximation. Here we explore how far one can go on describing the atmosphere at large as a two-dimensional CS model. Our approach is similar in spirit to other ideas applied to large-scale atmospheric dynamics and climate theory, such as \cite{Singh-ONeil}.

One of the salient features of the Earth's atmosphere at large scale is its essential two-dimensional nature --and this is also the case for many other terrestrial planets \cite{Zhang}. Roughly 70\% of our atmosphere's mass is contained in a layer ten kilometers thick extending over the surface some four thousand times larger which, to a very good approximation, is a two-dimensional sphere. The entire atmosphere can be viewed as a stack of thinner two-dimensional layers characterized by different specific thermodynamic and mechanical properties, interacting with the layers immediately above and below. Therefore, it could seem reasonable to describe the atmosphere as a system of two-dimensional fluids in which the effects of the vertical thickness are replaced by an interaction with the neighboring layers.  Such an approach differs considerably from the well-established and widely-used shallow water equation system, which does not completely neglect the vertical coordinate (and thus does not reduce the system into a 2-D model) but does consider a constant density profile. However, as described in detail later in this work, a clear interpretation of the comparison between the equation systems stemming from both approaches is possible.

The global mean state of the atmosphere is maintained by a statistical balance between sources and sinks of energy and momentum in the atmospheric circulation. Pioneer studies unveiled the similarity between the statistical characteristics of atmospheric motions and turbulence \cite{Sasamori-Melgarejo1978}. Hence, turbulence models have been used for the assessment of the predictability of weather and hence climate (see, e.g. \cite{Lorenz1969}). In this context, the very large horizontal-to-vertical scale ratio has been used to study the large-scale behavior of the atmosphere as a two-dimensional homogeneous isotropic turbulent flow \cite{Boer-Shepherd1983}. Several experiments have also used this approximation (see, e.g. \cite{Afanasyev-Wells2005}). This approach leads, for instance, to a better understanding of the internal variability in the atmosphere. The two-dimensional characterization corresponds to a more simplified approach to atmospheric motions than that of \cite{Charney1971}, where the quasi-two-dimensional nature of the atmosphere is described by means of the quasi-geostrophic model. That model, which considers constraints due to gravity, the Earth's rotation, and stratification, allows 1) the consideration of the effect of meridional baroclinicity (an external parameter) on the turbulence model through baroclinic unstable waves, and 2) an approximation to the power spectrum of vertical velocities \cite{Sasamori-Melgarejo1978}.

In the two-dimensional approximation, effects such as the vertical variation in air density, temperature and pressure, or the formation of the Hadley flow cells are ignored. In a description of the local behavior for regions of extensions comparable with the thickness of the atmosphere, however, the vertical displacements should not be ignored. Hence, the two-dimensional model cannot be expected to accurately describe local phenomena of great importance for weather forecasting, for example.

In this article we will consider the case of a single two-dimensional layer over a perfectly spherical rotating surface, driven by the inflow of energy coming from an external source representing the sun. The dynamical equations are derived from the approximation of the system as described by the first order CS equations, which is conceptually different from the standard SWE approach. The dynamical equations obtained from our analysis reflect many of the features of the standard approach, and in section \ref{sec:discussion}, the two approaches are compared. An important advantage of the equations we propose is that, in the small P\'eclet number approximation, they admit a family of analytic exact solutions parametrized by physical coefficients such as the average density, specific heat, thermal diffusion, emissivity and average energy flux from the sun. By varying these parameters, different global scenarios --and possibly different planets-- can be simulated with the aid of a code accessible through a link provided here.

The earth's atmosphere is a mixture dominated by air and water either in molecular dissolution, as suspended droplets, ice crystals, or as condensed water-air mixture \cite{Wallace}. In all these cases, the water content and temperature of the atmosphere determine the density and hence the inertia of the fluid as well as its thermodynamic features, such as the specific heat. In isolation, this fluid should obey conservation laws of mass and energy. The atmosphere, however, is not an isolated system but is constantly exchanging matter, energy and mechanical momentum with the planet's surface and with the exterior environment:

\textbf{Energy exchange:} The energy balance affects directly the temperature and pressure of the atmosphere and indirectly the flow patterns. Part of the radiation from the sun is directly absorbed by the atmosphere, part reaches the surface and is re-emitted and subsequently absorbed by the atmosphere, and the atmosphere also emits radiation to outer space. The distribution of land and oceans over the surface also plays an important role for these processes, in particular, due to spatial variations in albedo and heat capacity.

\textbf{Matter exchange:} The atmosphere changes its density by variations in temperature, pressure, altitude, and humidity. In particular, air parcels acquire moisture from evaporation and lose mass through precipitation. At the surface (e.g., 1000 hPa) and 20\textdegree C, the vapor content in 1 m$^3$ can change from 0 to 18 g (at saturation), which represents a density variation of about 1.5\%. These changes are affected by local mechanical and thermodynamic processes related in turn to surface properties. At high altitudes and colder temperatures, however, the air holds less water vapor and evaporation-precipitation is less significant than the exchange of air masses with the lower strata. Hence, in the case of an air layer with restricted vertical movements due to mechanical blocking (such as the surface) or thermodynamical conditions (e.g. stratification, as at the tropopause), local density fluctuations might be due to the vertical advection and turbulent entrainment.

\textbf{Mechanical interactions:} The surface interacts mechanically with the atmosphere due to the topographic features, affecting the local flow patterns. Again, this can be an important direct effect for the lower layers only. The Coriolis force, on the other hand, represents an important effect that should not be ignored at large scale.\\

In this simplified global description, a state of atmospheric fluid is characterized by a two-dimensional velocity field $\vec{v}(t,\vec{x})$, mass density $\rho(t,\vec{x})$, pressure $P(t,\vec{x})$ and temperature $T(t,\vec{x})$, where $t$ is time and $\vec{x}$ indicates a position on the 2-dimensional sphere. The dynamics of the system could be that of a compressible fluid moving under the influence of an energy source describing the Sun, and taking into account the Coriolis force due to the non-inertial reference frame attached to the rotating planet. We assume the two-dimensional fluid to be:

$\bullet$ An ideal gas, with uniform specific heat and compressibility, so that its internal energy is proportional to the temperature and to the pressure.

$\bullet$ Slightly dissipative, so that in the absence of external influences, the system would relax towards a static uniform equilibrium configuration. This dissipation takes the form of diffusion and damping of oscillations.

One can expect that this idealized model, where the interaction with the Earth's topography and local variations in albedo are initially neglected, could be a reasonable approximation to describe the layers of mid-troposphere to high troposphere at a global scale. A better approximation could be obtained by considering several interacting layers.

In this work, we are mainly interested in studying the steady state regime to which the system presumably relaxes. As shown in section \ref{sec:validation}, the initial conditions are eventually erased by dissipative effects. The same equations, however, could be used to explore short-time effects of large localized cataclysmic events such as a volcanic eruption or the impact of a meteorite.

\section{Chern-Simons single layer model}  
Consider the flow of a single atmospheric layer described by a three-component vector $A_\mu=(A_0, A_1, A_2)$, under the effect of external influences represented by the vector $j^\mu=(j^0,j^1, j^2)$. In this dynamical system, the field $A_\mu$ encodes the information about the velocity, density and pressure of a two-dimensional fluid in a three-dimensional spacetime $\mathcal{M}=\mathbb{R}\times S^2$, where a point in $\mathcal{M}$ has coordinates $(t, x^1, x^2)$. 

The Chern-Simons (CS) dynamics for a three-dimensional vector field takes the form of the system of equations (see Appendix \ref{app:CS})
\begin{align} \label{eq:CSeqs}
\partial_0 A_i - \partial_i A_0 = \epsilon_{ik} j^k\, ,\\
\partial_i A_k - \partial_k A_i = \epsilon_{ik} j^0\, .
\end{align}
Here $j^\mu=(j^0, j^i)$ corresponds to the interaction with external sources. These interactions include solar radiation --with its variations due to Earth rotation and orbital motion--, the heat emitted to outer space and also the exchanges of matter, whose integrated fluxes in a day should cancel out almost exactly. If the surface albedo varies --by e.g. land-use or land cover changes-- or the atmospheric emissivity is affected --by greenhouse gases, for example--, there can be long-term changes in the atmosphere's mean temperature.

Under the identification 
\begin{equation} \label{eq:ident}
 A_i = \rho v_i, \;\;\;\; A_0 = -P\, ,
\end{equation}
the above equations become\footnote{It can be directly checked that the dimensions (units) of all components of the one-form $A$ are the same: $[A_0dt]= [A_idx^i]= MT^{-1}$.}
\begin{align} \label{eq:CSeqs'-1}
\partial_t(\rho v_i) &= -\partial_i P + \epsilon_{ik}j^k\, ,\\ \label{eq:CSeqs'-2}
\hat{r}\cdot [\nabla \times (\rho \vec{v})] &= \partial_1(\rho v_2) -\partial_2(\rho v_1) = j^0\, .
\end{align}
The first equation is essentially a statement of Newton's second law in an instantaneous locally comoving frame, $d \vec{p}/dt= \vec{F}$, where $\vec{p}:= \rho \vec{v}$ is the momentum density. We neglect in this approximation the direct mechanical force exerted on the atmospheric fluid, such as the dragging due to the roughness of the Earth's surface. Hence, we assume $j^i$ as the force produced by the Coriolis effect (the centrifugal force can be reasonably neglected). By neglecting the external mechanical forces we are essentially ignoring molecular friction, a valid approximation for all motions in the Earth's atmosphere except for turbulent motions near the ground \cite{Holton}. Indeed, the contribution of frictional forces in the Earth's atmosphere  is 10$^{-9}$ times smaller than the Coriolis force (see, e.g. Table 2.1 in \cite{Holton}). This is an approximation that may be acceptable for the higher layers of the atmosphere but not for the lower strata, or for example much denser atmospheres. As we shall see, when properly written in the rest stationary frame of a grounded observer, \eqref{eq:CSeqs'-1} is essentially the Navier-Stokes equation (c.f. Eq.\eqref{eq:u} below).

Equation \eqref{eq:CSeqs'-2} relates the curl of the momentum density $\rho \vec{v}$ to some external influence $j^0$. The units of $j^0$ are $ML^{-2} T^{-1}$ (see Appendix B), which corresponds to the rate of change of density and therefore can be assumed to be proportional to the rate of change in time of $\rho$. Hence, \eqref{eq:CSeqs'-2} can be interpreted as a relation of the form $\hat{r}\cdot [\nabla \times \rho \vec{v}] \propto d\rho/dt$.
Clearly, the CS equations \eqref{eq:CSeqs'-1},\eqref{eq:CSeqs'-2} are not sufficient to describe the atmosphere. Precisely because it is an open dissipative system, it is necessary to include an additional equation to describe the thermodynamic changes produced by the exchanges of energy.

\subsection{Thermal energy balance} 
We will approximate the atmosphere as an ideal gas. This implies that the temperature and pressure could be related through an equation of state of the form
\begin{equation}\label{eq:stateeq} 
P(t,\vec{x})= R \rho(t,\vec{x}) T(t,\vec{x}) \, , 
\end{equation}
where $R$ is a constant, which depends on the nature and state of the fluid.\footnote{The dimensions of  $[R]$ are $L^2 T^{-2}[^o K]^{-1}$.} Substituting \eqref{eq:stateeq} in \eqref{eq:CSeqs'-1} relates the change in momentum of the fluid to the gradient of the thermal density $\tau\equiv \rho T$. Radiation from the sun as well as that reflected by the surface, and that emitted to outer space, produce changes in $\tau$. In addition, temperature may also change by diffusion, as heat flows from warmer to colder regions. Hence, we postulate that in a locally comoving frame, the energy density changes as
\begin{equation}\label{eq:heat-transfer}
\partial_t \tau - k\nabla^2 \tau = \frac{1}{c}E(t,\vec{x})\,,
\end{equation}
where $k$ is the diffusion coefficient, $c$ is the specific heat of the fluid and $E$ is the difference between energy received from the sun and that emitted to space per unit area and unit time. The specific heat of the atmosphere $c$ may depend on the phases in which the different gases in the mixture are found, but we will assume it to be constant.

Radiation also depends on the albedo and emissivity of surface features like oceans, forests, deserts, ice fields, etc. These features could be taken into account assuming $E$ to be a function of the position, 
\begin{equation}\label{eq:E}
E = \varphi(\vec{x}) \mathcal{E}(t, \vec{x})
\end{equation}
where $\mathcal{E}$ is the net energy arriving at the Earth and $\varphi$ is a dimensionless function that accounts for the local differences of surface. This local inhomogeniety becomes less relevant for atmospheric layers well above the surface. Moreover, for the sake of simplicity, in this study we will neglect this effect by using $\varphi=1$. If the function $E(t,\vec{x})$ is known, equation \eqref{eq:heat-transfer} can be integrated using the standard methods. Energy, as described in \eqref{eq:E}, will be used as a forcing term in section \ref{sec:sphericalcoords}. In addition, section \ref{sec:energybalance} describes the energy balance in detail.

\subsection{Matter density balance} 
A given layer of the atmosphere gains and loses matter by exchange with the neighboring layers, turning the continuity equation into a non-conservation relation. The density also changes as the fluid expands and contracts by the changes in pressure, which can be modelled as a diffusion process that tends to homogenize the density in the absence of other external drivers. Hence, we postulate
\begin{equation} \label{eq:continuity}
\partial_t \rho + \nabla \cdot(\rho \vec{v})=\sigma \nabla^2 \rho + \beta j^0 \, ,
\end{equation}
where we have added on the right hand side the dissipative term with diffusion coefficient $\sigma$, plus the term $\beta j^0$ representing changes in the matter content which, according to \eqref{eq:CSeqs'-2}, is related to the curl of the momentum density. Thus, the time evolution of $\rho$ becomes
\begin{equation} \label{eq:rho-0}
(\partial_t  -\sigma\nabla^2) \rho = \beta \hat{r}\cdot [\nabla \times \vec{p}]- \nabla \cdot \vec{p} \,. 
\end{equation}
Here $\beta$ is a dimensionless coefficient that determines the gain or loss of matter by the atmosphere due to vertical flow in a rotational wind pattern. A puzzling feature of this contribution is that, unlike the remaining terms in \eqref{eq:rho-0}, it is parity odd. It changes sign under reflections in latitude\footnote{Here latitude ($-\pi/2\leq \theta\leq \pi/2$), and longitude ($0\leq \phi\leq 2\pi$), are the coordinates on the sphere.} ($\theta \to -\theta$), or longitude ($\phi \to -\phi$), while the other terms in \eqref{eq:rho-0} are invariant. The planet's rotation, however, breaks the east-west symmetry but it does not affect the north-south symmetry. Hence, in order to respect the symmetry of the system, we are led to postulate $\beta$ to be an odd function of the latitude, $\beta(-\theta)=-\beta(\theta)$. A natural option is 
\begin{equation}
    \beta(\theta) = \beta_0 \sin \theta\,,
\end{equation}
although more general expressions such as $\beta(\theta)=\sum_n \beta_n \sin \theta \cos^n\theta$ could also be useful to describe different atmospheric layers. 

The density of a substance is defined as its mass per unit volume, but here we are interested in a two-dimensional density, i.e., expressed in units of mass per unit area. This value is obtained using the exponential approximation form for air density as a function of altitude, given by
\begin{equation}\label{eq:rho(h)}
    \rho(h) = \rho_{atm}\, e^{-h/H}\,,
\end{equation}
where $\rho_{atm}$ is the standard atmospheric volumetric density value at sea level, and $H \approx 10$km is the height scale of the exponential fall \cite{Wallace}. Integrating equation \eqref{eq:rho(h)} from the point of interest (e.g., the top of the troposphere) to infinity gives the two-dimensional density value.

\subsection{Momentum equations}  
Equation \eqref{eq:CSeqs'-1} could be solved for the momentum density $\vec{p}$, but this equation represents an idealized situation. In fact, in this equation, $\vec{p}$ could be shifted by a constant $\vec{p}_0$ that would not decay even if $j^i=0$, which is clearly unphysical. Even worse, as we shall see, when the system is periodically driven --by the daily influence of the sun and the Coriolis frequency--, this equation has unbounded resonant solutions for $\vec{p}$. In a realistic scenario, however, there is always some damping due to viscosity or other forms of dissipation of mechanical energy, which prevent the divergent resonant behavior.

\subsubsection{Rotation} 
So far, the equations of motion for the atmosphere \eqref{eq:CSeqs'-1}, \eqref{eq:heat-transfer}  and \eqref{eq:rho-0} describe a situation on a static planet. In order to consider the effects of rotation, the change from a static to a rotating frame brings in the Coriolis and centrifugal forces, which result from the change of variables
\begin{equation}
\theta \to \theta \;,\; \phi \to \phi + \omega t \;,
\end{equation}
where $\omega=(1/86,400)Hz$ is the rotational frequency of the Earth. Thus, following \cite{Arfken} and neglecting the centrifugal force proportional to $\omega^2$, the equation for $\vec{p}$ becomes
\begin{equation} \label{eq:vecu}
\left(\partial_t +\; 2\vec{\omega} \times\;\right) \vec{p}|_{\perp} =  -R \nabla \tau\;,
\end{equation}
where the symbol $|_{\perp}$ indicates the projection of the vector on the surface. The cross product between $\vec{\omega}$ and $\vec{p}$ depends on the latitude $\theta$, so that $\vec{\omega} \times\ \vec{p}|_{\perp}=(\omega \sin \theta )\hat{r}\times \vec{p}$.

\subsubsection{Damping}  
We close this section setting up the equations for $\vec{p}$ in a form that can be easily integrated and that avoids an unphysical resonant behavior. Differentiating \eqref{eq:vecu} with respect to time gives
\begin{equation}\label{harm-osc}
(\partial_t^2 + \nu^2 (\theta))\vec{p} = -R(\partial_t - \nu \hat{r}\times) \nabla \tau\;,  
\end{equation}
where $\nu (\theta)\equiv 2\omega \sin \theta$ is the Coriolis frequency. In polar coordinates this equation reads
\begin{align}
\label{eq:uddot1}
\left[ \partial^2_t + \nu^2(\theta) \right] p^\theta = f^\theta(t,\theta,\phi)\,, \\\label{eq:uddot2}
\left[ \partial^2_t + \nu^2(\theta) \right] p^\phi = f^\phi(t,\theta,\phi)\,,
\end{align}
where the driving forces are
\begin{align}
f^\theta(t,\theta,\phi) &= -\frac{R}{r_0} \left[\partial_t\partial_\theta \tau - \frac{\nu(\theta)}{\cos{\theta}} \partial_\phi \tau \right] \,, \\
f^\phi(t,\theta,\phi) &= -\frac{R}{r_0} \left[ \frac{1}{\cos{\theta}}\partial_t\partial_\phi \tau + \nu(\theta)\partial_\theta \tau \right] \,,
\end{align}
with $r_0$ being the Earth's radius.
Equations (\ref{eq:uddot1},\ref{eq:uddot2}) describe a two-dimensional {\it undamped} harmonic oscillator of resonant frequency $\nu(\theta)$, driven by the external time-dependent forces $f^\theta$ and $f^\phi$. If $\tau$ is periodic with a frequency $\omega_0\leq 2 \omega$, there is always a certain latitude $\theta$ for which $\omega_0=2 \omega |\sin \theta|$, producing an unbounded resonance. Of course, an oscillator system with no damping is an idealized situation. A more realistic scenario should include a dissipative term $\eta p^i$ on the left hand side of \eqref{eq:vecu}. Differentiating once more the new equations with respect to time yields the damped harmonic oscillator equations
\begin{align}
\label{eq:uddot1-damp}
\left[ \partial^2_t + 2\eta\,\partial_t + (\eta^2+\nu^2(\theta)) \right] p^\theta = F^\theta(t,\theta,\phi)\,, \\\label{eq:uddot2-damp}
\left[ \partial^2_t + 2\eta\,\partial_t + (\eta^2+\nu^2(\theta)) \right] p^\phi = F^\phi(t,\theta,\phi)\,,
\end{align}
where $2\eta$ is the damping coefficient, the resonant frequency has been shifted from $\nu$ to $\sqrt{\nu^2 + \eta^2}$, and the driving forces are now given by
\begin{align} \label{vecF1}
F^\theta(t,\theta,\phi) &= -\frac{R}{r_0} \left[(\partial_t + \eta)\partial_\theta\tau - \frac{\nu(\theta)}{\cos{\theta}} \partial_\phi \tau \right] \,, \\ \label{vecF2}
F^\phi(t,\theta,\phi) &= -\frac{R}{r_0} \left[ \frac{1}{\cos{\theta}}(\partial_t + \eta)\partial_\phi \tau + \nu(\theta)\partial_\theta \tau \right] \,.
\end{align}

\subsection{Advection}  
The equations for $\tau$ and $\vec{p}$ above give the time evolution in a comoving frame. For an external observer it is more convenient to relate the field variables as functions of spatial coordinates fixed relative to the surface of the planet. In order to translate the results to the reference frame of the surface, we should replace $\partial_t$ by the material derivative, $\partial_t + \vec{v}\cdot \nabla$ in \eqref{eq:heat-transfer} and \eqref{eq:vecu}, which takes into account the changes in the physical variables at a fixed point of the surface due to the drift of the fluid.

Thus, the set of equations one should solve to determine $\tau$, $\vec{p}$ and $\rho$ is\footnote{Here all vectors except $\hat{r}$ are tangent to the sphere and $\nabla \tau = r_0^{-1}(\hat{\theta} \partial_\theta \tau + \hat{\phi}[1/\cos \theta]\partial_\phi \tau)$, etc.}
\begin{eqnarray}\label{eq:tau} 
\left(\partial_t - k \nabla^2 + \vec{v}\cdot \nabla \right) \tau &=& \frac{1}{c}E(t,\vec{x}) \;, \\
\label{eq:u}
\left(\partial_t +\eta + \nu \, \hat{r}\times + \,\vec{v}\cdot \nabla \;\right) \vec{p} &=&  -R \nabla \tau\;, \\
\label{eq:ro}
\left(\partial_t - \sigma \nabla^2 \; \right)\rho &=& \beta(\theta)\, \hat{r}\cdot( \nabla \times \vec{p}) - \nabla \cdot \vec{p}  \;.
\end{eqnarray}
This set of coupled differential equations is linear for $\tau$ and $\rho$, but the advective term introduces a nonlinearity into the equation for $\vec{p}$. Note that if the advective term can be dropped\footnote{Which could be the case for small P\'eclet or Reynolds number.} from \eqref{eq:tau} and \eqref{eq:u}, the system becomes linear and decoupled: The first equation is solved for $\tau$ as the convolution product of $E$ and the diffusive Green function. Then, \eqref{eq:u} determines the momentum, which in turn determines $\rho$.

\section{Linearized fluid equations} \label{eq:fluideqs}  
If the advective terms cannot be dropped, there are alternative approaches to integrate these equations:
\begin{enumerate}[(i)]
\item Numerical integration of the time evolution from an initial configuration,
\item Perturbative series expansion around the linearized solution, or
\item Replacing the velocity field $\vec{v}$ in the left hand side of (\ref{eq:tau},\ref{eq:u}) by its average $\vec{v}_0$.
\end{enumerate}
The first approach is standard but requires a judicious choice of the initial conditions and careful handling of instabilities due to the nonlinear and possibly chaotic behavior.
The second alternative can be useful if the nonlinear effects can be regarded as small corrections. Since in the absence of advection, the system (\ref{eq:tau}-\ref{eq:ro}) is linear and can be solved by standard methods, the effect of nonlinearities can be added as small corrections to the linear problem. 

Dropping the advection term could be a valid approximation if the characteristic wind speeds are small compared to the other relevant velocities of the system. The Rossby number is defined as the ratio of the sizes of the advective and Coriolis terms. Here, as in the geostrophic approximation, ignoring advective terms is motivated by a scale analysis of the horizontal momentum equation for large-scale flow in the Earth's atmosphere. In this case, the Rossby number is small ($\sim$10$^{-1}$), and thus rotation effects dominate over advection.

Similarly, if diffusion is significantly greater than advection at the relevant regime, the advection term can also dropped from the equation for $\tau$, which would be the case if the corresponding P\'eclet number is small.

The third option consists in approximating the velocity field by the time average $\vec{v}_0$ in the expectation that this average can account for the advection, capturing the behavioral pattern of the evolution over long periods compared with the daily cycles.

In the remaining of this work, we concentrate on the solution of the linearized case, which will also serve to validate the model in an important limit. By using this simplification, our model will be aimed at describing large-scale flow far from the equator but not short-lived sub-synoptic phenomena \cite{Vallis}.

\subsection{Spherical coordinates} 
\label{sec:sphericalcoords}
We write the equations in spherical coordinates ($\theta$,$\phi$), where $\theta = -\pi/2$ is the south pole and $\theta =\pi/2$ the north pole, and $0\leq \phi \leq 2\pi$ is the azimuthal angle.
\begin{align}
\label{eq:dot(tau)r}
 \left(\partial_t - \frac{k}{r_0^2\cos{\theta}}\left[\partial_\theta (\cos{\theta}\,\partial_\theta\; )+\frac{1}{\cos{\theta}}\partial_\phi^2\right] \right)\tau &= \frac{1}{c} E(t,\theta,\phi) \,, \\
\label{eq:dot(u1)r}
 (\partial_t + \eta\,) p^{\theta} + 2\omega\sin{\theta}\,p^{\phi} &= -\frac{R}{r_0}\partial_\theta \tau \,, \\
\label{eq:dot(u2)r}
 (\partial_t + \eta\;) p^{\phi} - 2\omega\sin{\theta}\,p^{\theta} &= -\frac{R}{r_0\cos{\theta}}\partial_\phi \tau \,,  \\ 
 \label{eq:dot(rho)r}
 \left(\partial_t -\frac{\sigma}{r_0^2\cos{\theta}}\left[\partial_\theta (\cos{\theta}\,\partial_\theta \; ) + 
    \frac{1}{\cos{\theta}}\partial_\phi^2\;\right]\right)\rho &= 
    \!\begin{aligned}[t]
    &-\frac{\partial_{\theta}(\cos{\theta}\,p^{\theta}) + \partial_{\phi}p^{\phi}}{r_0\cos{\theta}}\\   
    & + \beta(\theta) \frac{\partial_{\theta}(\cos{\theta}\,p^{\phi})-\partial_{\phi}p^{\theta}}{r_0\cos{\theta}}\,.
    \end{aligned}
\end{align} 
\subsection{General solution of the linear problem} 
\label{sec:generalsol}
In what follows, we present the analytic solution of the linear problem defined by equations (\ref{eq:dot(tau)r} -- \ref{eq:dot(rho)r}). 

\subsubsection{Thermal density $\tau(t,\vec{x})$:} 
Equation \eqref{eq:dot(tau)r} can be integrated using the Green function for the diffusion equation,
\begin{align}\nonumber
\tau(t,\theta,\phi) = \frac{1}{c}\int_{0}^{2\pi}d\phi' \int_{-\pi/2}^{\pi/2} d\theta' \int_{-\infty}^{\infty} dt' G_{k}(t,t';\theta,\theta';\phi,\phi') & E(t',\theta',\phi')\cos{\theta'}\,\\
& + \tau_h(t,\theta,\phi) \,, \label{eq:Intrgal-tau}
\end{align}
where $\tau_h(\theta, \phi)$ is a solution of the homogeneous problem that matches the appropriate initial conditions. The Green function \cite{Arfken} is given by 
\begin{align} \label{eq:Gd}
G_{k}(t,t';\theta,\theta';\phi,\phi') &= \sum_{l=0}^{\infty}\sum_{m=-l}^{l}g_{k,l}(t,t')\,Y_{l m}^*(\theta',\phi')\,Y_{l m}(\theta,\phi)\,,\\
g_{k,l}(t,t') &= \,\Theta(t-t')\,e^{-k\frac{l(l+1)}{r_0^2}(t-t')}\,,
\end{align}
where $\Theta(t-t')$ is the step function and $Y_{l m}(\theta,\phi)$ are the spherical harmonics.
\subsubsection{Momentum density $\vec{p}(t,\vec{x})$} 
Again, the solution can be found through the Green function for a damped harmonic oscillator,
 \begin{equation} \label{eq:ui}
 p^i(t,\theta,\phi) = \int_{-\infty}^{\infty}G_{\eta}(t-t')F^i(t',\theta,\phi)\,dt' \; + \; p^i_h(t, \theta, \phi), \qquad i = \theta\,, \; \phi\;,
 \end{equation} 
 where $F^i$ is given in (\ref{vecF1}, \ref{vecF2}), and
\begin{equation}
 G_{\eta}(t-t') = \frac{e^{-\eta\,t}\sin(\nu(\theta)\,t)}{\nu(\theta)}\,, \;\; \mbox{with}\;\; \nu(\theta)= 2\omega \sin{\theta}\;,
 \end{equation}
and ($p^\theta_h\,, p^\phi_h$) are solutions of the homogeneous problem that meet the initial conditions. For $F^i=0$, the momentum density $\vec{p}$ decays exponentially, vanishing for large $t$ and therefore, we take $\vec{p}_h=0$. The integration of \eqref{eq:dot(u1)r}, \eqref{eq:dot(u2)r} is greatly simplified by the fact that the angular dependence is parametric.\\

\subsubsection{Matter density $\rho(t,\vec{x})$:} 
Finally, equation \eqref{eq:dot(rho)r} is integrated along the same lines as \eqref{eq:dot(tau)r}, 
\begin{align} \label{eq:Intrgal-rho} \nonumber
\rho(t,\theta,\phi) = \int_{0}^{2\pi} \int_{-\pi/2}^{\pi/2}  \int_{-\infty}^{\infty} G_{\sigma}(t,t';\theta,\theta';\phi,\phi') & F(t',\theta',\phi')\cos{\theta'}\,dt'd\theta'd\phi'\\
& + \rho_h(t,\theta,\phi) \,, 
\end{align}
where $G_{\sigma}(t,t';\theta,\theta';\phi,\phi')$ has the same expression as \eqref{eq:Gd} with $k$ substituted by $\sigma$ and $\rho_h(\theta, \phi)$ is the solution to the homogeneous problem with appropriate initial conditions which can be taken as a constant. The source function in convolution with the Green function is now given by
\begin{equation} \label{eq:source_rho}
F(t,\theta,\phi)= \beta(\theta)\frac{\partial_{\theta}(\cos{\theta}\, p^{\phi})-\partial_{\phi}p^{\theta}}{r_0\cos{\theta}} - \frac{\partial_{\theta}(\cos{\theta}\, p^{\theta}) + \partial_{\phi}p^{\phi}}{r_0\cos{\theta}}\,.
\end{equation}

Let us summarize what we have so far. Equations (\ref{eq:dot(tau)r} -- \ref{eq:dot(rho)r}) have been analytically solved for $\tau$, $\vec{p}$ and $\rho$. These solutions depend on the energy source $E(t,\theta,\phi)$ and the parameters specific heat ($c$), diffusion coefficients for heat and matter ($k, \sigma$), ideal gas constant ($R$), curl coefficient ($\beta$), viscous damping ($\eta$), rotation frequency ($\omega$) and Earth's radius ($r_0$). The values and ranges of these parameters are discussed in Appendix \ref{app:units}.

\section{Validation}  
\label{sec:validation}

As an illustration of the integration rationale and to validate the model, we discuss an idealized situation where the surface features -oceans, continents, ice covered regions, mountains- are ignored, corresponding to a constant $\varphi$ in \eqref{eq:E}; the energy provided by the Sun is constant and uniformly distributed, and the fluid moves as a single layer governed by (\ref{eq:dot(tau)r}-\ref{eq:dot(rho)r}). We will see how the model presented here can be integrated and will explore solutions \eqref{eq:Intrgal-tau}, \eqref{eq:ui}, \eqref{eq:Intrgal-rho} for a particular form of $E$ in an homogeneous planet. 

In order for the dynamical equations to yield a unique solution, it is necessary and sufficient to specify appropriate boundary conditions in space and time for the diffusion equations \eqref{eq:dot(tau)r} and \eqref{eq:dot(rho)r}, and initial conditions for \eqref{eq:dot(u1)r} and \eqref{eq:dot(u2)r} \cite{Courant-Hilbert}. Since the two-sphere has no boundary, boundary conditions for the Laplacian are not required (it is replaced by single-valuedness). On the other hand, the idea of initial conditions for a planet's atmosphere is rather artificial. Hence, we substitute the initial conditions by other requirements on the solution which have a more natural interpretation and which are sufficient to yield a unique solution. These are the equilibrium conditions for $\tau$, $\vec{p}$ and $\rho$.

For $\tau$, we expect that the radiation absorbed and re-emitted to space balance almost exactly in a period $2\pi/\omega$, producing an equilibrium energy density $\tau_0$. In the case of $\vec{p}$, one expects that for a constant $\tau$ --so that the right hand side of \eqref{eq:dot(u1)r} and \eqref{eq:dot(u2)r} vanish--, $\vec{p}$ would be damped, approaching zero exponentially with time. Finally, for $\rho$ we expect that for a pattern of time-independent momentum density $\vec{p}_0$, defined by the time-average of $\vec{p}(t,\theta. \phi)$, the matter density will reach a time-independent equilibrium configuration $\rho_0(\theta, \phi)$.\\

\subsection{Energy balance} 
\label{sec:energybalance}
\begin{figure}[h!]
\centering
\includegraphics[width=6cm]{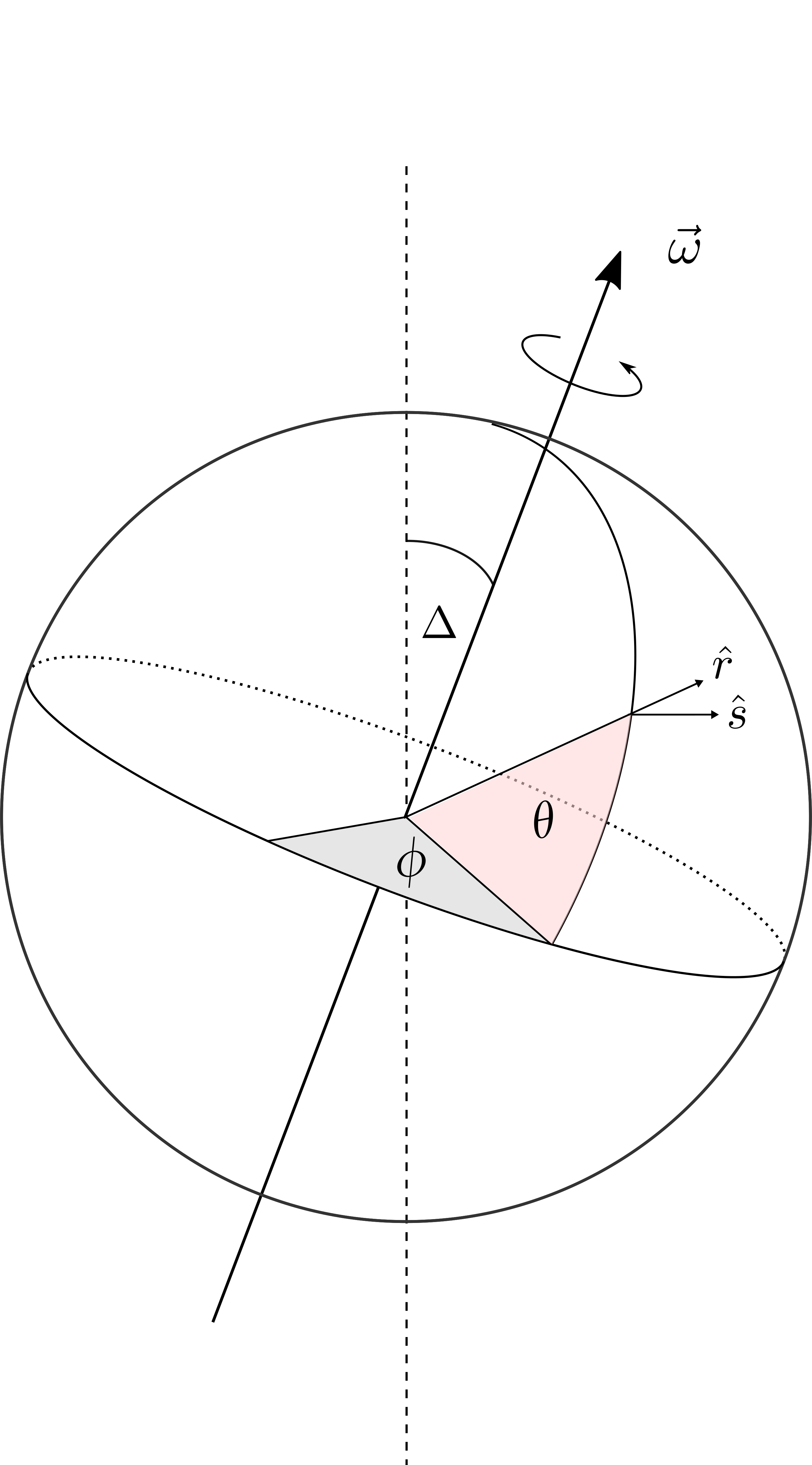}
\caption{Tilt angle $\Delta$ between the rotation axis and the normal to the orbital plane (dotted line).}	
\label{fig:coordinates}	
\end{figure}

The energy function $E$ on the right side of \eqref{eq:dot(tau)r} accounts for the net energy entering the atmosphere, driving the whole system. If no other significant energy sources are present, one can assume $E$ to be the result of the absorbed and emitted energies
\begin{align} \label{eq:Ein-Eout}
E(t,\theta,\phi) = E_{in} (t,\theta,\phi) - E_{out} (t,\theta,\phi)\,.
\end{align}
The first term is the energy entering the atmosphere as radiation either directly from the sun or reflected from the surface. Assuming a constant and uniform albedo, this incoming energy can be assumed to be proportional to the radiation from the sun during the day, which takes the form
\begin{align} \label{eq:Ein}
E_{in}(t,\theta,\phi) & = \mathcal{E}_{0}\,\hat{r} \cdot \hat{s} \; \Theta(\hat{r} \cdot \hat{s})\,,
\end{align}
where $\mathcal{E}_{0}$ is some fraction of the solar radiative flux density and $\Theta$ is the Heaviside's step function, $\hat{r}$ is the unit vector in the radial direction and $\hat{s}$ is the unit vector in the direction of the Sun (see Figure \ref{fig:coordinates}). If the tilt angle between the rotation axis and the normal to the orbital plane is $\Delta$, then 
\begin{equation} \label{r-dot-s}
\hat{r} \cdot \hat{s} = \cos{\Delta(t)} \cos{\theta} \sin{(\phi +\omega t)} + \sin{\Delta(t)} \sin{\theta} \;,
\end{equation}
where $\Delta(t)=\Delta_0 \cos{\Omega t}$ and $\Omega$ is the orbital frequency (for the Earth, $\Omega=\omega/365$, $\Delta_0=23.4\degree$). In this work we will ignore this seasonal effect in order to simplify the discussion. 
For the energy radiated away, we postulate an emission rate to outer space similar to a black-body radiation form $E_{out} = \lambda \tau(t,\theta,\phi)$, where $\lambda$ is an emissivity coefficient.\footnote{If the emission were not directly proportional to $\tau$, like in the Stefan-Boltzmann law, $E_{out}=\sigma(\tau/\rho)^4$, the formulas below would remain basically unchanged with redefined values for $\lambda$ and mean thermal density $\tau_0$.}\\

The net energy gained by the system is the difference
\begin{equation}
E(t,\theta,\phi) = E_{in} (t,\theta,\phi) - \lambda \tau(t,\theta,\phi)\;.
\end{equation}
Hence, neglecting the advective term, equation \eqref{eq:tau} for $\tau$ becomes
\begin{equation} \label{dif-lambda}
\left(\partial_t - k \nabla^2 + \frac{\lambda}{c} \right) \tau = \frac{1}{c}E_{in}(t,\vec{x}) \;,
\end{equation}
which can be solved using the same Green function in \eqref{eq:Gd} for the case without emission:
\begin{equation} \label{eq:tau-lambda}
\tau(t,\vec{x}) = \frac{1}{c}\int G_k(t,t';\vec{x},\vec{x}')\,e^{-\frac{\lambda}{c}(t-t')}\, E_{in}(t',\vec{x}') dt'd\vec{x}'  + \tau_h \;.
\end{equation}
Here $\tau_h$ is the homogeneous solution of \eqref{dif-lambda}, which reads 
\begin{equation} \label{44}
\tau_h = \sum_{l,m}\tau_{lm} e^{-Q_l t}Y_{lm}(\theta, \phi)\;,\;\mbox{where}\;\; Q_l=k\frac{l(l+1)}{r_0^2}+\frac{\lambda}{c}\,.
\end{equation}
For finite $\lambda$, $\tau_h$ vanishes as $t \to \infty$. However, the monopole mode $Q_0$ is highly sensitive to $\lambda$. In particular, for very small emissivity ($\lambda<<1$), $\tau_h$ would vanish very slowly for $t \to \infty$, and what is more crucial, the exponential suppression in \eqref{eq:tau-lambda} would be insufficient to prevent $\tau$ from reaching an arbitrary large value.
\subsection{Integration} 
\label{sec:integration}
Substituting the expression for $E_{in}$ from \eqref{eq:Ein} in \eqref{eq:tau-lambda} gives
\begin{equation}\label{eq:Gen-tau}
    \tau(t, \theta, \phi) = \frac{\mathcal{E}_{0}}{c}\sum_{l,m} N_{lm}^2 \,D_{lm}(t)\, P_l^m(\sin\theta)\, e^{im\phi}\, +\, \sum_{l,m} \tau_{lm} Y_{lm}(\theta,\phi)\,e^{-Q_l t}\;,
\end{equation}
where
\begin{align}
& N_{lm} = \sqrt{\frac{(2l+1)(l-m)!}{4\pi (l+m)!}}\\
\label{55}
& D_{lm}(t) = \int_{-\pi/2}^{\pi/2}\cos{\theta'}\;  P_l^m(\sin{\theta'})\; H_{lm}(t,\theta') \;d\theta'\\ \label{56}
& H_{lm}(t,\theta') = \int_{0}^{t} C_m(t', \theta')\,  e^{-Q_l(t-t')}dt'\;\, \\
\label{57}
& C_m (t', \theta')= \int_0^{2\pi} e^{-im\phi'} (\hat{r}\cdot \hat{s}) \Theta(\hat{r}\cdot \hat{s}) d\phi'\\
& \tau_{lm} = \int_{-\pi/2}^{\pi/2} \int_{0}^{2\pi} \tau_i\, N_{lm}\,P_l^m(\sin\theta)\, e^{im\phi} \cos{\theta} d\theta\, d\phi
\end{align}
The last integral vanishes for $\tau_i=constant$ for all $(m,l)\neq (0,0)$ and therefore the last term in \eqref{eq:Gen-tau} is just $\tau_i e^{-\lambda t/c}$. The scalar product $\hat{r}\cdot \hat{s}$ involves $t$, $\theta$ and $\phi$, but still \eqref{57} and \eqref{56} can be explicitly integrated. The integral in \eqref{55} is not elementary except for the special case $\Delta=0$, which is discussed below.

In \eqref{56} we see that for small $\lambda$, $H_{00}$ could grow arbitrarily large with $t$, producing an arbitrarily large finite contribution to $\tau(t, \theta, \phi)$.

\subsubsection{Transient and steady state components} 

The solution described by \eqref{eq:Gen-tau} corresponds to the evolution of the atmosphere from an initial state $\tau(0,\theta,\phi)=\sum_{l,m} \tau_{lm} Y_{lm}(\theta,\phi)$. The general behavior is that of an oscillation with the frequency of the driving force $E_{in}$ around an average that evolves from the initial value to a new equilibrium state. After a sufficiently long time (longer than $2\pi/\omega$), the steady state is established and the only trace of the transient phenomenon is to be found in the phase difference between the driver $E_{in}(t,\theta, \phi)$ and the response $\tau(t,\theta, \phi)$. That phase shift and the amplitude of the oscillation depend on the parameters of the differential equation ($c,k,\lambda,\mathcal{E}_0,\omega$) and on the latitude ($\theta$).

\subsubsection{Global space and time average $\tau_0$} 
The existence of an equilibrium mean temperature for the atmosphere ($T_0=\tau_0/\rho_s$ where $\rho_s$ is the average density of the two-dimensional atmosphere) means that the time average of the absorbed and emitted radiations must balance, otherwise the energy would increase --or decrease-- until a new equilibrium average temperature is reached. The condition on $\lambda$, $\tau_0$ and $\mathcal{E}_0$ for which this equilibrium is reached, can be obtained from the equation
\begin{equation}\label{eq:int(E)}
\int_{S^2} d\Omega \int_0^{2\pi/\omega} E_{in}(t, \theta,\phi)dt  =  \int_{S^2} d\Omega \int_0^{2\pi/\omega} E_{out}(t, \theta,\phi)dt \,.
\end{equation}
Direct substitution of \eqref{eq:Ein} (for $\Delta=0$) on the left hand side of \eqref{eq:int(E)} gives $2\pi^2 r_0^2\mathcal{E}_{0}/\omega$. Using the form $\tau=\tau_0+\Delta\tau$, where $\Delta\tau$ is the fluctuation around the average $\tau_0$ and vanishes integrated over a period, it is easy to see that
\begin{equation} \label{eq:tau(lambda)}
\tau_0 = \frac{\mathcal{E}_0}{4\lambda}.
\end{equation}
This equation establishes the average global temperature as 
\begin{equation}
T_0 = \frac{\mathcal{E}_0}{4\lambda \rho_s}\,,
\end{equation}
which shows that for a given energy flow $\mathcal{E}_0$ and average atmospheric density $\rho_s$, the equilibrium average temperature is inversely proportional to the emissivity $\lambda$.

\subsubsection{Explicit solution ($\Delta=0$)} 
Following the steps outlined above, we construct the solution for $(\tau, \vec{p}, \rho)$ which, in the special case $\Delta=0$, is completely reducible to elementary integrals. \\ 

{\bf Thermal density} \\
The solution \eqref{eq:Intrgal-tau} reads
\begin{equation} \label{eq:tau''}
\tau(t,\theta,\phi)= \frac{\mathcal{E}_0}{c}\sum_{l,m} \frac{s_m\,e^{-i\delta_{lm}} (e^{i m \omega t} - e^{-Q_l t})}{\sqrt{Q_l^2 + m^2 \omega^2}}\,K_{lm}\, N_{lm} \,Y_{lm}(\theta,\phi) +  \tau_i e^{-\lambda t/c}\;,
\end{equation}
where $\sin{\delta_{lm}}= m\omega/\sqrt{Q_l^2 + m^2 \omega^2}$,  $K_{lm}=\int_{-\pi/2}^{\pi/2} d\theta' \cos^2{\theta'} \,P_l^m(\theta')$, and $s_m$ is $\pm(\pi/2i)$ for $m=\pm 1$, $0$ for odd $m\neq \pm 1$, and $-2/(m^2-1)$ for even $m$. For $t>>2\pi/\omega$, the transient terms can be dropped; then, using \eqref{eq:tau(lambda)} and keeping only the dominant terms in the harmonic expansion ($l=0,1,2$), we find for $\tau$ (see Appendix C):
\begin{align} \nonumber
\tau(t,\theta,\phi) &= \tau_0 + \\  \label{eq:tau''} & \frac{\mathcal{E}_0}{2c }\left[ \frac{\cos{\theta} \sin(\phi +\omega t -\delta_{11})}{\sqrt{\omega^2 + Q_1 ^2}} + \frac{5 (3\cos 2\theta -1)}{64  Q_2} - \frac{15 \cos^2\theta \cos(2[\phi+\omega t] - \delta_{22})}{32\sqrt{4\omega^2 + Q_2^2}}\right].
\end{align}
The different components show a periodic evolution with the same frequency as $E_{in}$, but phase-shifted by $\delta_{lm}$ relative to this driver. The phase shifts $\delta_{lm}$ are combined effects of rotation ($m \omega $), diffusion ($k/r_0^2$), and heat loss due to emission ($\lambda/c$). The contributions for $l=3, 4,...$ are smaller and typically involve higher harmonics in $\theta$ and additional phase shifts. The first term on the right hand side of (55) is the global average; the second is the first harmonic contribution ($l=1$) that follows the dipole form of the energy source, with a phase shift $\delta_{11}$. The last two terms come from the second harmonics $(l=2)$, where the third is the long-term effect of the predominant equatorial warming, smeared uniformly in the longitudinal direction ($m=0$), and the fourth is a second harmonic term ($m=2$) with a further phase shift $\delta_{22}$.\\

{\bf Momentum density}\\
Using the steady state solution for $\tau$ above, the steady state solution for $\vec{p}$ is found to be
\setlength{\jot}{12pt}
\begin{align}\label{eq:u_sol}
p^{\theta}(t,\theta,\phi) =-\frac{R\mathcal{E}_0}{r_0c} \sum_{l,m} \frac{J_{lm}\left([im\omega+\eta] \partial_{\theta} P_l^m(\sin{\theta}) \,-\, 2 im\omega \tan{\theta}\,P_l^m(\sin{\theta}) \right)\, e^{i m (\phi+\omega t)}}{[4 w^2 \sin ^2{\theta} + (im w + \eta )^2]}\nonumber \\
p^{\phi}(t,\theta,\phi) = -\frac{R\mathcal{E}_0}{r_0c} \sum_{l,m} \frac{J_{lm}\left(\left[\frac{im\omega +\eta}{\cos{\theta}}\right]im P_l^m(\sin{\theta}) \,+\, 2\omega \sin{\theta} \partial_{\theta} P_l^m(\sin{\theta}) \right) e^{i m (\phi+\omega t)}}{[4 w^2 \sin ^2{\theta} + (im w + \eta)^2]}\,, 
\end{align}
where $J_{lm} = N_{lm}^2 K_{lm} s_m/(Q_l + i m \omega)$. Note that changing $m\to -m$ is equivalent to complex conjugation. Hence, symmetric sums over $m$ always produce a real result. The first few non-zero harmonic terms are given in Appendix \ref{app:integration}. 

The average velocity can be estimated as $\vec{v}_0 = \vec{p}_0/\rho_0$, where $\vec{p}_0$ can be found averaging \eqref{eq:u_sol} over a period $2\pi/\omega$, 
\begin{equation} \label{u0}
\vec{p}_0(\theta) = -\frac{R \mathcal{E}_0}{r_0c} \frac{\sum_l  J_l\, \partial_\theta P_l(\sin \theta)}{(2\omega \sin \theta)^2 + \eta^2} (\eta\,, \,2\omega \sin \theta)\,,
\end{equation}
where $J_l=(2l+1)K_l/(2\pi Q_l)$, with $K_l=\int_{-1}^{1} dx \sqrt{1-x^2}P_l(x)$, and $\rho_0$ is the global average [c.f., Eq. \eqref{ro-equil}]. The average momentum density corresponds to the $m=0$ term in the sum \eqref{eq:u_sol} and it is also the average in $\phi$. The average $\vec{p}_0$ represents a steady momentum flow resulting from the rotation of the planet under the influence of the external heat source (the sun). The momentum density field can be viewed as the sum of the average $\vec{p}_0(\theta)$ and the time-dependent fluctuation coming from the components with $m\neq 0$ in \eqref{eq:u_sol}, so that
\begin{equation} \label{split-u}
\vec{p}(t,\theta, \phi) = \vec{p}_0(\theta) + \sum_{l, m\neq 0} \vec{p}_{lm} (\theta)\, e^{im(\phi + \omega t)}.
\end{equation} \\

{\bf Matter density}\\
Using the solution \eqref{eq:u_sol} and the source function $F(t,\theta, \phi)$ given by \eqref{eq:source_rho}, the convolution \eqref{eq:Intrgal-rho} yields the steady state form for the density (see Appendix \ref{app:integration})

\begin{equation} 
\label{eq:rho_sol}
\rho(t,\theta,\phi) = \int d\Omega' \int_{0}^{t} dt'  \sum_{l=0}^{\infty}\sum_{m=-l}^{l} Y_{l m}(\theta,\phi) Y_{l m}^*(\theta',\phi') \,e^{-\frac{\sigma l (l+1)}{r_0^2}(t-t')} F(t',\theta',\phi')\,,
\end{equation}
where $d\Omega'=\cos \theta'd\theta'd\phi'$. Since the driver $F$ is a linear operator acting on $\vec{p}$, \eqref{split-u} induces the splitting
\begin{equation}
F(t,\theta, \phi) = F_0(\theta) + \sum_{l, m\neq 0} F_{lm} (\theta)\, e^{im(\phi + \omega t)}\,, 
\end{equation}
where $F_0$ is given by \eqref{eq:source_rho} evaluated for $\vec{p}=\vec{p}_0$. The density $\rho$ satisfies a linear equation and we will assume that it oscillates around an equilibrium distribution $\rho_0$, defined as the one produced by the steady flow of momentum $\vec{p}_0$. In other words,
\begin{equation} \label{ro-equil}
- \sigma \nabla^2 \rho_0 = F_0(\theta).
\end{equation}
The time independence and axial symmetry of the problem imply that $\rho_0=\rho_0(\theta)$ and therefore we expect the solution to take the form
\begin{equation}
\label{eq:rho_decomposed}
\rho(t,\theta,\phi) = \rho_0(\theta) + \sum_{l, m\neq 0} \rho_{lm} (t,\theta, \phi)\, ,
\end{equation}
where $\rho_{lm} (t,\theta, \phi)$ is given by expression \eqref{eq:rho_sol} with $F(t,\theta, \phi)$ substitued by $F_{lm} (\theta)\, e^{im(\phi + \omega t)}$, and with $m=0$ omitted in the sum. There is an indeterminate additive constant $\bar{\rho}$ in $\rho_0$, fixed by the condition
\begin{equation}
  R^2\int (\rho_0(\theta)+\bar{\rho}) d\Omega = M\,,
\end{equation}
where $M$ is the mass of the entire atmosphere. Details for the first harmonics in the density are given in Appendix \ref{app:integration}.
\subsubsection{Extension for $\Delta \neq 0$} 
Taking into account the tilt of the rotation axis relative to the normal to the ecliptic plane is challenging. Following the steps outlined at the beginning of this section, it is still possible integrate $t'$, $\phi'$ and $\theta'$ for fixed nonzero $\Delta$, although this time it is less trivial. The result, however, is unsatisfactory as it would describe a planet receiving radiation from a source at a fixed angle, as in permanent solstice. This would produce a warm region around the pole in constant summer and a cold region around the other.

An exact solution is in principle still possible for the source given by (\ref{eq:Ein}, \ref{r-dot-s}). The solution would involve periodic functions of frequencies $\omega$, $\Omega$, as well as their difference and their sum. If $\omega >> \Omega$, as in the case of our planet, the solution could be reasonably approximated by a high frequency function ($f(\phi+\omega t$)) modulated by a low frequency factor ($h(\Omega t)$). The discussion of such solution, however, lies beyond the scope of this paper, whose purpose is to explore the general features of the model.

An idea of the annual pattern can be obtained by substituting $\theta$ by $\theta +\Delta(t)$ in the solution for $\Delta=0$. For small amplitude $\Delta_0$ this would be a good approximation in a nearly stationary regime with $\Omega<<\omega$.

\subsection{Simulations} 
\begin{figure}[h!]
\centering
\includegraphics[width=16.2cm]{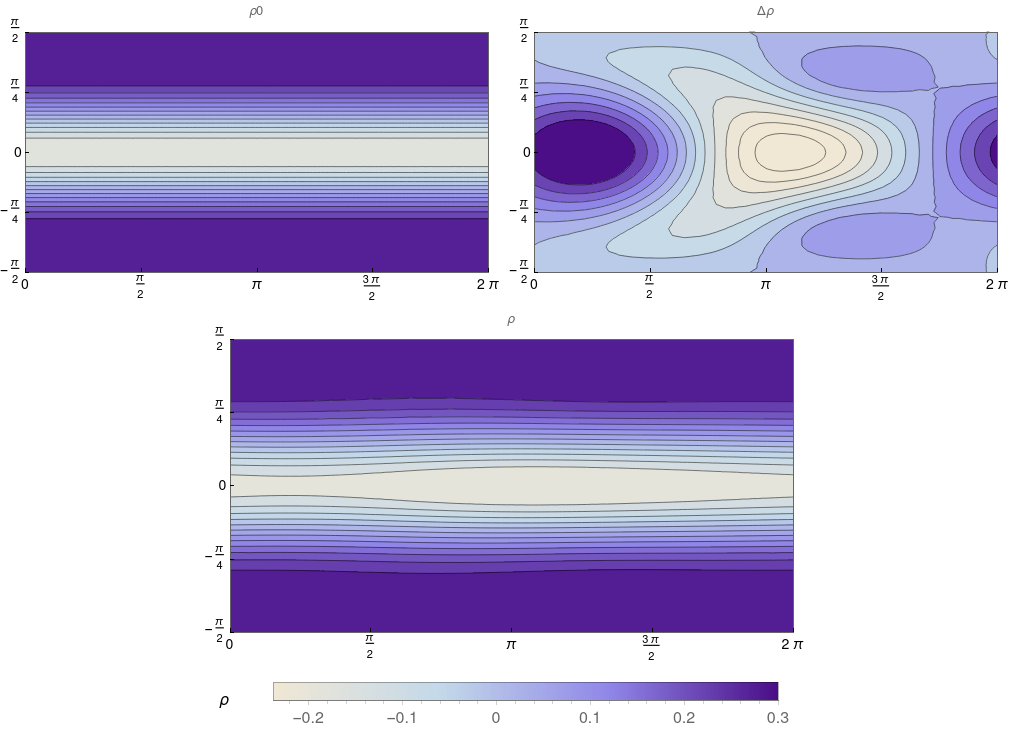}
\caption{Density distribution simulated for a non-tilted planet using the default parameter values given in Table \ref{tab:params}. Panels show the time-and-axial independent component $\rho_0(\theta)$ without the additive constant $\bar{\rho}$ (top left), the oscillation component $\Delta\rho(t,\theta,\phi)$ given by the right hand second term in \eqref{eq:rho_decomposed} (top right), and the complete solution $\rho$ given by the sum of the previous components (bottom). The scale for the variation $\Delta\rho$ is 1/14 of the one shown in the bar legend at the bottom.}
\label{fig:rho}	
\end{figure}

\begin{figure}[h]
\centering
\subfigure[]{
\includegraphics[width=14.0cm]{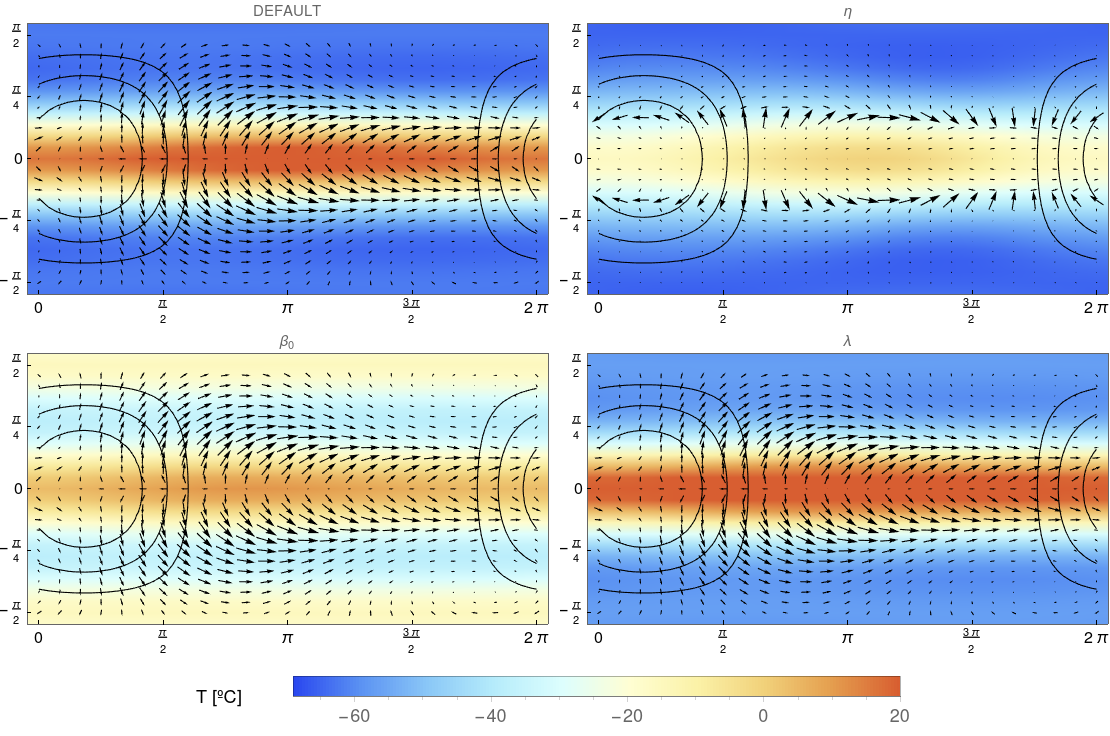}}
\subfigure[]{
\includegraphics[width=14.0cm]{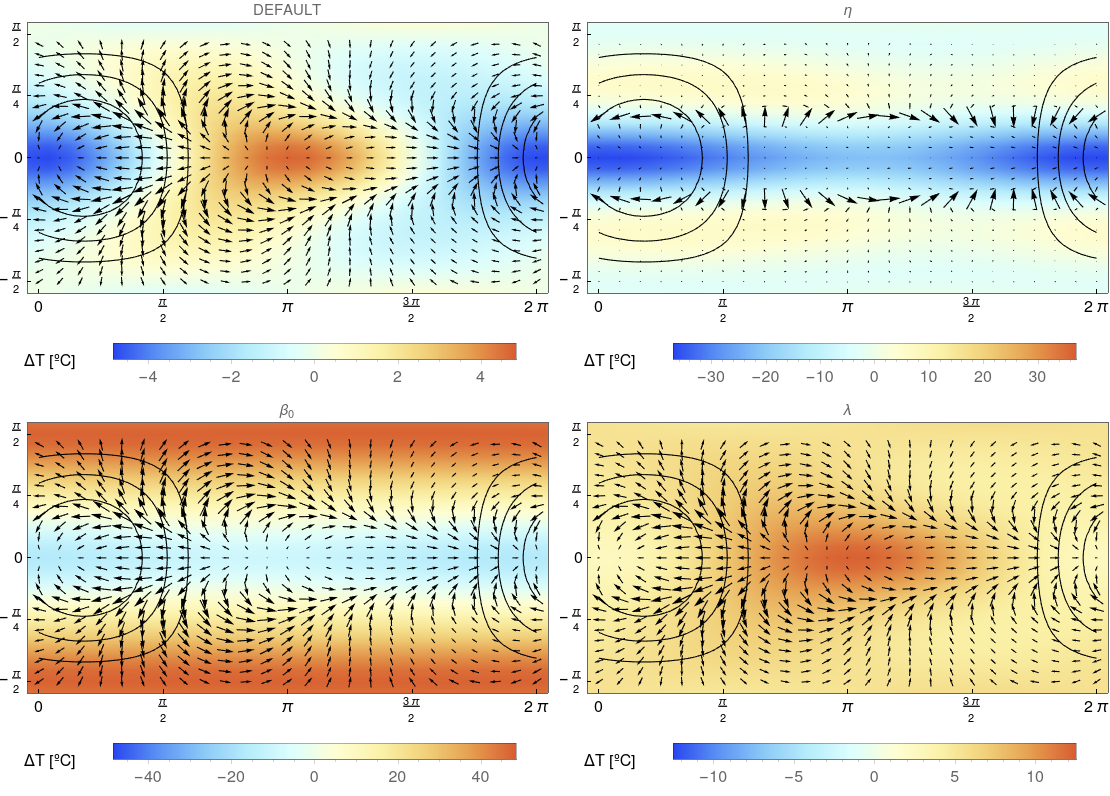}}
\caption{Absolute (a) and anomalies (b) velocity and temperature fields in a non-tilted planet. For each four-panel chart upper left are the fields for the default parameter values given in Table \ref{tab:params} ($\eta=0.5$ , $\beta_0=-2.0$ , $\lambda=0.004$). One of these parameters is changed in each of the other figures, keeping the default values: upper right $\eta = 0.05$, bottom left $\beta_0=-0.1$ and bottom right $\lambda=0.039$. The maximum magnitude of the velocity is near 4 m/s for all the panels except for when $\eta$ is varied for which the maximum is approx. 7 m/s. Contours in the background show the position of sun's radiation.}		
\label{fig:v&T}	
\end{figure}

The plots of $\rho$, $T$ and $\vec{v}$  for an untilted planet are presented in Figures \ref{fig:rho} and \ref{fig:v&T}. We show a ``default" (upper-left panel in Figure \ref{fig:v&T}) along with three ``perturbed" simulations: for the latter ones, one single parameter is modified in each case. As can be seen en Figure \ref{fig:rho}, for the set of default parameters chosen in the simulations (see Table \ref{tab:params}, Appendix \ref{app:units}), $\rho$ turns out to be quite uniform, varying by less than 2\% in time and 29\% in space, each of its entire range. This spatial variation is comparable to that of a dry standard atmosphere at surface: ~20\% considering a temperature range between -20$\degree$C and 30$\degree$C \cite{StandardAtmosphere}). Thus, the velocity field in this case follows closely the same patterns as the momentum field $\vec{p}$. For the simulations displayed $\beta(\theta)=-\beta_0 \sin\theta$.\\
The effect of varying the parameters ($\mathcal{E}_0$, $\lambda$, $c$, $k$, $R$, $\eta$, $\sigma$, $\beta_0$) in the equations can be explored using the Python interactive code in \cite{App}.

%

\clearpage
\begin{table}[h!]
\renewcommand{\arraystretch}{1.4}
\centering
\begin{tabular}{|c|c|c|c|c|}
\hline
Latitude ($\degree$)   &       Default     &       $\eta = 0.05$       & $\beta_0 = -0.1$      & $\lambda$ = 0.0039 \\ \hline
0 & 17.1 -- 26.7   & -15.4 -- 0.2   & 5.1 -- 12.2    & 24.6 -- 34.4 \\ \hline
15 & 0.7 -- 8.6     & -21.6 -- -9.3  & -5.7 -- 0.3    & 7.7 -- 15.7     \\ \hline
30 & -30.9 -- -25.9 & -35.7 -- -29.5 & -25.8 -- -21.9 & -24.7 -- -19.6 \\ \hline
45 & -54.5 -- -51.3 & -49.4 -- -47.4 & -37.1 -- -34.9 & -48.9 -- -45.6 \\ \hline
60 & -63.9 -- -61.9 & -58.5 -- -57.7 & -33.7 -- -32.5 & -58.5 -- -56.5 \\ \hline
75 & -63.7 -- -62.7 & -63.1 -- -62.5 & -21.4 -- -20.9 & -58.3 -- -57.3 \\ \hline
90 & -62.2      & -65.1          & -14.5          & -56.7            \\ \hline
\end{tabular}
\caption{Ranges of temperature (in $\degree$C) for the default values, $\eta=0.5$, $\beta_0=-2.0$ and $\lambda = 0.004$ (second column) for different latitudes. Columns 3 to 5 correspond to three scenarios in which those parameters take different values, one at a time (cases shown in Figure \ref{fig:v&T}).}
\label{tab:temperature_ranges}
\end{table}

Figure \ref{fig:v&T} exhibits two sets of panels. Each simulation is identified by the title on its top. For each one, panel set a) shows the actual fields of every variable, whereas panel set b) shows the anomalies of each field with respect to the corresponding zonal mean calculated from the ``default" simulation.\\

From Figure \ref{fig:v&T}, panel b, the ``default" simulation suggests a Matsuno-Gill response \cite{Matsuno,Gill}, with a Rossby wave in the form of a pair of anticyclones, each one located poleward and westward of the tropical warm anomaly. To the East, a Kelvin wave is apparent. This resemblance will be discussed below. Such warm anomaly reaches around +4$\degree$C and lags the maximum Sun's radiation by approximately 10 hours. This spatial structure is present in all but one simulation: that with reduced damping, $\eta$.\\

Table \ref{tab:temperature_ranges} refers to values computed to inter-compare each simulation. We see that: reducing $\eta$ to 1/10 of its default value produces a cooling by 30$\degree$C of the equatorial belt and a 3$\degree$C warming of the polar regions; decreasing 20 times the absolute value of $\beta_0$ cools the equator by 13$\degree$C and warms the polar regions by up to 47$\degree$C. Finally, a modest reduction by 2.5$\%$ in $\lambda$ raises the average temperature between 5.5$\degree$C and 7$\degree$C globally.\\

\subsection*{ Comments}
The following list is meant to emphasise some features and limitations of the model itself, give an interpretation of the conservation of mass equation and further discuss the model's sensitivity to some parameters as illustrated in the simulations.\\

\noindent
{\bf Synoptic limits}\\
In the Earth's atmosphere, the advective terms are of a similar order of magnitude as the local acceleration \cite{Holton} and should not be neglected in a realistic weather model for the Earth. Hence, the absence of synoptic systems, particularly in mid- and high-latitudes, is a consequence of neglecting the advective non-linear terms in the momentum equation \eqref{eq:CSeqs'-1}.\\

\noindent
{\bf Eddy statistics}\\
As pointed out by \cite{Holton-UL}, for the simulation of an Earth-like upper-level atmosphere (i.e., with geostrophy constraints on the zonal mean zonal wind and temperature fields and given distributions of radiative heating and zonal wind),  two-dimensional models are not able to resolve realistically the mean meridional motion and the mean zonal flow tendency without considering eddy statistics of heat and momentum sources. Nevertheless, diffusive damping has been used to model eddy statistics, a strategy which is not valid in regions where the horizontal eddy momentum fluxes concentrate momentum rather than diffuse it, e.g. in the polar night jet.\\

\noindent
{\bf Turbulent dynamics}\\
Regarding mean turbulent motions within the Earth's boundary layer, terms associated with molecular diffusion are smaller by a factor $10^{-7}$ than the rest of the terms involved in the momentum equation \cite{Stull}. This aspect further highlights the relevance of including non-linear terms in equation \eqref{eq:CSeqs'-1} for a realistic representation of turbulent dynamics near the surface.\\

\begin{figure}[h!]
\centering
\includegraphics[width=15cm]{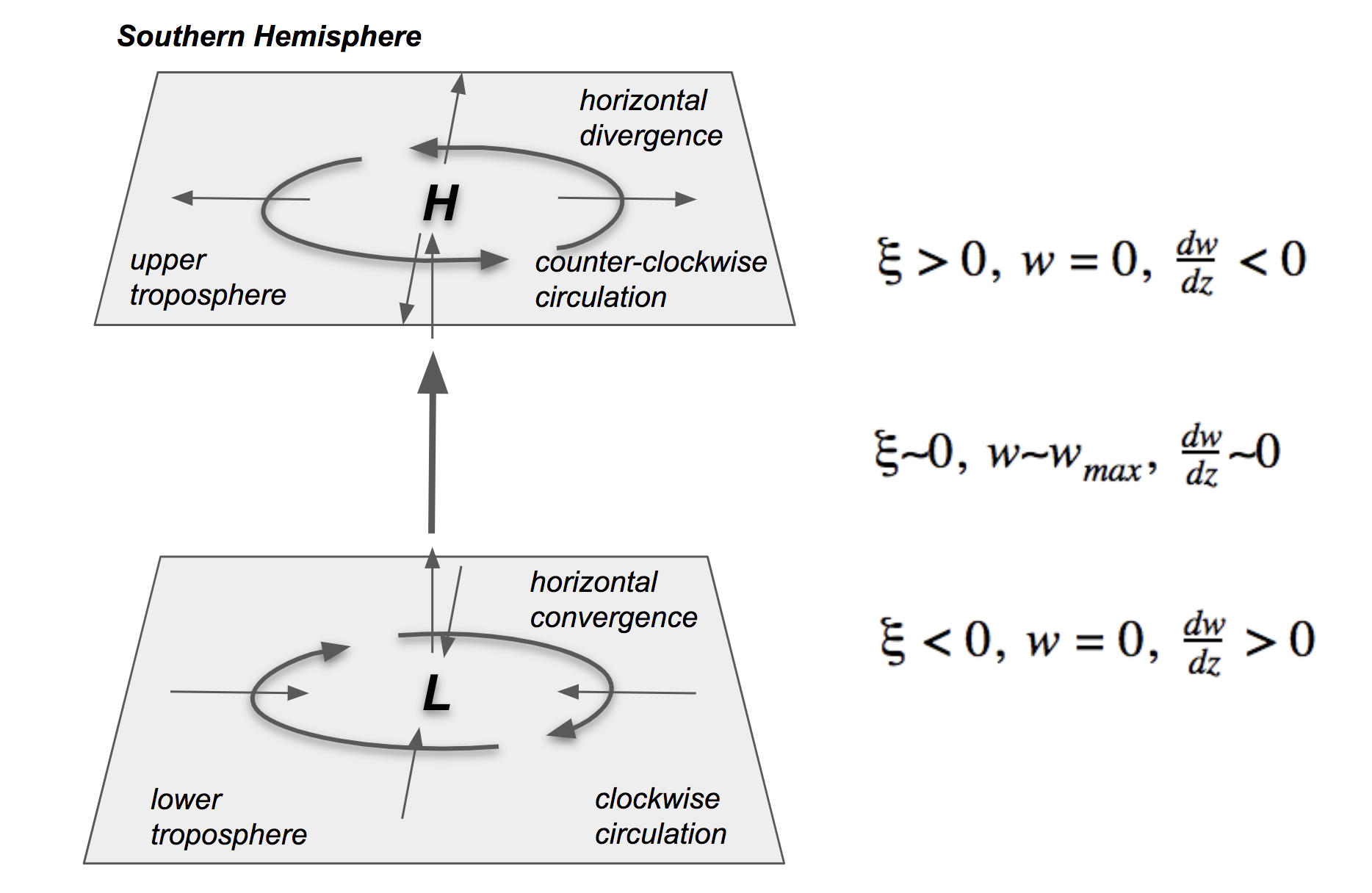}
\caption{Schematic relationship between horizontal convergence/divergence and vertical velocity at different levels of the troposphere. L and H denote low and high pressure areas, respectively. The illustrated case corresponds to the Southern Hemisphere, where $\beta(\theta)>0$. In the the Northern Hemisphere, $\beta(\theta)<0$, the horizontal circulation reverses, and thus $\xi$ exhibits opposite signs. }
\label{fig:SH}	
\end{figure}

\noindent
{\bf Matter density balance and momentum curl}\\
The simulated momentum density fields exhibit divergent (convergent) patterns at regions of maxima (minima) of thermal density, which, in turn, follow the energy input distribution by a known phase. Because our model is restricted to 2 dimensions and no vertical motions are allowed, as it would be expected within the quasi-geostrophic model for the conservation of angular momentum, the matter density balance to induce a curl, as seen in eq. \eqref{eq:ro}. This phenomenon can be understood in a pseudo-3D context: a straight-forward interpretation of eq. \eqref{eq:ro} might be achieved by setting $\rho=constant$. In that case,
\begin{equation} \label{ro:cte}
\nabla \cdot \vec{v} = \beta(\theta)\, \hat{r}\cdot( \nabla \times \vec{v})
\end{equation}
\noindent
Considering the 3D continuity equation, 
\begin{equation} \label{continuity3d}
\frac{du}{dx} + \frac{dv}{dy} + \frac{dw}{dz} = 0\,,
\end{equation}
\noindent
where $\vec{v} = (u,v,w)$ is the 3D wind vector. Hence, the left-hand side term of \eqref{ro:cte} might be interpreted as the vertical gradient of the pseudo-vertical velocity: 
\begin{equation} \label{pseudovertical}
-\frac{dw}{dz} = \beta(\theta)\, \hat{r}\cdot( \nabla \times \vec{v}) = \beta(\theta)\, \xi
\end{equation}
\noindent
Thus, the vertical gradient of the pseudo-vertical velocity equals $-\beta(\theta)$ times the horizontal component of the relative vorticity, $\xi=\hat{r}\cdot( \nabla \times \vec{v})$. From eq. \eqref{continuity3d} and Fig. \ref{fig:SH}, we appreciate the similarity to the Ekman pumping phenomenon. If $\beta(\theta)$ is an odd function and positive in the Southern Hemisphere, we identify this phenomenon as the vertical motions induced in the atmosphere by horizontal divergence or convergence at a certain level due to changes in $\xi$, as schematized in Figure \ref{fig:SH}. This might be associated with e.g. the development of planetary waves. By choosing $\beta(\theta)=\beta_0 \sin \theta$ (an adimensional parameter), we set the maximum of the the relative vorticity at the poles and annul this term at the equator. In this way, we expect relatively high values of relative vorticity at mid to high latitudes, which is a realistic feature of the Earth's atmosphere.\\

\clearpage

\noindent
{\bf Parametric sensitivity}\\
Figure 3 illustrates the sensitivity of the model to changes in the parameters $\eta$, $\lambda$, and $\beta_0$. The default values given in Table 1, produce the velocity and temperature patterns in the upper left panel. Reducing the damping to 1/10 of the default value produces a pronounced resonance at $30^o$  latitude, dispersing the thermal energy from the equatorial region towards the poles and also spreading it along the equatorial belt. Compared to the Earth, in a denser atmosphere such as that of Venus, or less dense such as the one in Mars, $\eta$ would be much larger or smaller, respectively. This would give rise to reduced or enhanced resonance in the tropical regions, respectively.

The model is highly sensitive to changes in $\lambda$, as seen in the lower right panel: reducing $\lambda$ by just 2.5\% from its default value produces a warming of the entire planet by around six degrees. In particular, in the Earth's atmosphere an increase in $CO_2$ is expected to lower atmospheric emissivity expressed as a reduction in  $\lambda$.

Reducing $|\beta_0|$ to 1/20 of its default value induces a significant reduction of density in the polar regions, which is responsible of the increase in temperature at high latitudes. According to \eqref{ro:cte}, the divergence (and hence the pseudo-vertical velocity) is proportional to $|\beta_0|$. Hence, reducing its value implies a less dynamic (more static) scenario, with ``poorly-mixed" density distribution. 

A further interpretation of $\beta$ might be derived from the analogy between our eq. \eqref{pseudovertical}, used for the case of air with constant density --where we introduced a pseudo-vertical velocity to our 2D theoretical framework-- and the widely-used quasi-geostrophic vorticity equation for the free atmosphere:

\begin{equation} \label{vorticity_g}
\dfrac{\mathrm{D}\xi_g}{\mathrm{D}t}  = f_0 \frac{\partial w}{\partial z}\,,
\end{equation}
\noindent
where the label $g$ refers to the geostrophic approximation. As in \cite{Vallis}, we are now interested in the case of an atmosphere that can be modelled as a homogeneous single layer of thickness $H$ underlaid by an Ekman layer of thickness $d<<H$; the latter representing the effects of friction and stress near the surface. Neglecting the vertical velocity at the top of the free atmosphere ($z=H+d$), eq. (5.218) of \cite{Vallis} shows that:
\begin{equation} \label{vorticity_g_ekman}
\dfrac{\mathrm{D}\xi_g}{\mathrm{D}t}  = -\frac{f_0 d}{2H}\xi_g\ \,.
\end{equation}
\noindent
In \cite{Vallis}, this is identified as a ``linear drag" acting on the interior flow due to the Ekman layer, with $r=\frac{f_0 d}{2H}$ the drag coefficient. In our particular case, eqs. \eqref{pseudovertical}, \eqref{vorticity_g}, and \eqref{vorticity_g_ekman} lead us to identify $\beta=\frac{d}{2H}$. In other words, $\beta$ is proportional to such drag coefficient. Hence, a more (less) viscous, and thus deeper (shallower) Ekman layer will be represented in our 2D model by a relatively large (small) value of $\beta$.

\section{Discussion}  
\label{sec:discussion}
The simplified model presented here describes a two-dimensional atmosphere on a uniform spherical planet rotating on an axis not necessarily orthogonal to the ecliptic. The atmosphere is described by a fluid of temperature $T(t,\vec{x})$, velocity $\vec{v}(t,\vec{x})$ and density $\rho(t,\vec{x})$, satisfying the equations (\ref{eq:tau}-\ref{eq:ro}). Those equations are similar --but not identical-- to those obtained by E. N. Lorenz in 1948 \cite{Lorenz-1948}. Following Lorenz, the current standard approach to atmosphere dynamics usually includes the Navier-Stokes equation supplemented by the matter balance and thermodynamic equations, as for example in the text of G. K. Vallis  \cite{Vallis}. In those two cases the system is described by the three-dimensional velocity, density and thermal energy of the fluid. The Shallow Water Equations, on the other hand, are derived from depth-integrating the Navier-Stokes equations, and describe a system where the horizontal length scale is much greater than the vertical length scale. The question then is under what assumptions the two-dimensional model described here by (\ref{eq:tau}-\ref{eq:ro}) can be related to these standard equations.

\subsection{Relation to 3D dynamics}  
In order to make the analysis more transparent we will write each equation in the corresponding models. Whenever confusion can arise, we distinguish three-dimensional and two-dimensional quantities by a sub-index, $(...)_3$ and $(...)_2$, respectively.\\
\subsubsection{Comparison with the standard models} 
{\bf I. Velocity}
\begin{eqnarray}
\mbox{Lorenz/Vallis:} \quad \qquad \qquad \rho D\vec{v} - \nu\rho \nabla^2 \vec{v} &=& -\nabla P + \vec{F} \;,\quad  \label{eq:Lorenz-v}\\
\mbox{SWE:} \quad \quad \qquad \qquad \qquad D\vec{u} \qquad &=& -g\nabla h + \vec{f} \times \vec{u} \;,\quad  \label{eq:swe-v}\\
\mbox{Us:} \; \quad \quad  \qquad \qquad \qquad D \vec{p}  +\eta \, \vec{p} &=& -\nabla P + \vec{F}_C \;.\quad  \label{eq:Us-v}
\end{eqnarray}
where $D=\partial_t + \vec{v}\cdot \nabla$ is the material derivative\footnote{Since we are interested in the horizontal flows, we assume $(D)_3=\partial_t + (\vec{v}\cdot \nabla)_3 \approx \partial_t + (\vec{v}\cdot \nabla)_2 =(D)_2$.}, $\vec{F}$ is the external mechanical force density, $\vec{u}$ is the horizontal velocity, $\vec{f}=2\,\omega\sin{\theta}\hat{r}$ is the Coriolis parameter, $g$ is the gravitational acceleration, $h$ is the free surface height (and the thickness in the flat bottom case) of the thin layer composing the shallow water system, and $\vec{F}_C$ is the Coriolis force.\\

{\bf II. Matter conservation}
\begin{eqnarray}
\mbox{Lorenz/Vallis:} \quad \qquad \qquad \qquad D\rho +\rho \nabla \cdot \vec{v} &=& 0\;,  \label{eq:Lorenz-r}\\
\mbox{SWE:} \quad \quad \qquad \qquad Dh + h\nabla \cdot \vec{u} &=& 0 \;,  \label{eq:swe-r}\\
 \qquad  \quad \mbox{Us:}  \quad \qquad \;  D\rho + \rho \nabla \cdot \vec{v} -\sigma \nabla^2 \rho &=& \beta(\theta) \hat{r}\cdot \nabla \times \vec{p} \,.  \label{eq:Us-r}
\end{eqnarray}

{\bf III. Thermodynamics}
\begin{eqnarray}
\mbox{Lorenz:} \qquad \qquad \qquad \qquad \qquad P &=& (Const) \rho^\lambda\;, \label{eq:Lorenz-T}\\
\mbox{Vallis:} \quad \qquad \qquad \; D I + P D w &=& \dot{Q}\;, \label{eq:Vallis-T}\\
 \mbox{Us:}\qquad \quad \qquad \;\; D \tau - k\nabla^2\tau &=& \frac{1}{c}E \;, \label{eq:Us-T}
\end{eqnarray}
where $I=c_vT$ is the specific internal energy, $w=1/\rho$ is the specific volume of the fluid and $\dot{Q}$ is a form of thermodynamic dissipation. 

\begin{enumerate}[(i)] 
\item Apart from the obvious fact that we have suppressed the vertical dependence, \eqref{eq:Us-v} can be compared with Lorenz's equations as follows. Combining \eqref{eq:Lorenz-v} and \eqref{eq:Lorenz-r} yields $D\vec{p} + (\nabla \cdot \vec{v}) \vec{p} - \nu \rho \nabla^2 \vec{v}= -\nabla P + \vec{F}$, which has the form \eqref{eq:Us-v}, where the external mechanical force corresponds to the Coriolis force $\vec{F}=-\vec{\omega}\times\vec{p}$, and the damping term in \eqref{eq:Us-v} is identified with a combination of two terms,
\[
\eta \vec{p} = (\nabla \cdot \vec{v})_3 \vec{p} - \nu \rho (\nabla^2 \vec{v})_3.
\]
In this interpretation, the damping parameter $\eta$ is an effective coefficient accounting for the fact that mechanical energy is transformed into heat due to compression ($\nabla \cdot \vec{v}$) and diffusion ($\nabla^2 \vec{v}$).

On the other hand, comparing Eqs. \eqref{eq:swe-v} and \eqref{eq:Us-v} we observe that for constant $\rho$ and vanishing $\eta$, our equation reduces to the SWE, where the horizontal pressure gradient $\nabla P$ can be identified with $\rho g\nabla h$, thus recovering the hydrostatic balance assumed in the SWE approach. Therefore, our equations extend the SWE to the case in which the density is not constant, allowing for dissipative damping as well.

\item Equations \eqref{eq:Lorenz-r}  and \eqref{eq:swe-r} represent the continuity equation, where the latter holds in the case of constant $\rho$ and therefore the matter contained in a column is just proportional to its height. Our expression \eqref{eq:Us-r} is the two-dimensional rendition of \eqref{eq:Lorenz-r}, which also allows for diffusion and influx of matter represented by the vertical component of the curl, $\beta(\theta)\hat{r}\cdot\nabla \times \vec{p}$. Hence, also in this case the SWE can be seen as a particular case of our equation obtained in the limit $\sigma \rightarrow 0, \beta\rightarrow 0$. (Eq. \eqref{eq:Lorenz-r} is also obtained in this limit)

\item Combining Eqs. \eqref{eq:Vallis-T} and \eqref{eq:Lorenz-r}, and assuming the ideal gas relation between pressure and temperature, one obtains
\begin{equation} \label{heating}
D \tau + \gamma (\nabla \cdot \vec{v})_3 \tau  = \frac{\rho}{c_v} \dot{Q}\,,
\end{equation}
where $\gamma=c_p / c_v$. Matching with \eqref{eq:Us-T}, this requires identifying
\begin{equation} \label{matching tau}
\frac{\rho}{c_v} \dot{Q} = \frac{E}{c_p} + \gamma (\nabla \cdot \vec{v})_3 \tau + k (\nabla^2 \tau)_2\,,
\end{equation}
which means that the flow of thermal energy into the system given by $\rho\dot{Q}/c_v$ is distributed into the net energy heating the atmosphere $E/c_p$ plus dilution and diffusion.

\end{enumerate}

\subsubsection{Comparison with the Matsuno-Gill model} 

The steady state equation \eqref{eq:Us-v} resembles the dimensionless momentum shallow-water equations proposed by Matsuno for the atmospheric response to diabatic forcing confined to the tropics \cite{Matsuno}. Moreover, the modal version of the continuity equation proposed by Gill \cite{Gill} is similar to the steady state of equation \eqref{eq:Us-r}, in the limit of negligible diffusion ($\sigma \to 0$). This suggests that this kind of atmospheric phenomena could be approachable by our model as well.

The model of Matsuno and Gill that describes the dynamics of the atmosphere in the equatorial belt is described (in dimensionless form) by the equations
\begin{align}\label{MG1}
\partial_t \vec{p} -\frac{y}{2}\hat{r}\times \vec{p} = - \nabla P\,, \\
\label{MG2}
\partial_t P + \nabla \cdot \vec{p}= -Q\,.
\end{align}

Here $(y/2)\hat{r}\times \vec{p}$ represents the Coriolis force near the equator, where $4\omega \sin \theta \approx y$ in re-scaled dimensionless units, $\vec{p}$ is the horizontal velocity for constant $\rho$ in dimensionless units and $Q$ is proportional to the heating rate. Thus \eqref{MG1} is essentially \eqref{eq:Us-v} if we neglect advection and damping in our equations (cf. \eqref{eq:u}). In the ideal gas approximation $P=R \tau$, and therefore \eqref{MG2} should be compared with \eqref{eq:Us-T}. Again, dropping the advection term these two equations agree if we identify $-Q=\frac{R}{c_p} E + \nabla \cdot \vec{p} +R k \nabla^2 \tau $, which means that heating is distributed into absorbed energy, adiabatic expansion and diffusion. 

\subsection{Multi-layer model} 
Equations (\ref{eq:tau}-\ref{eq:ro}) describe the dynamics of one isolated two-dimensional atmosphere driven by an external energy source. In a more realistic approach, the atmosphere can be conceived as a multilayer system in which each two-dimensional component interacts with the neighboring layers above and below. This could be modelled by a system of equations that generalizes (\ref{eq:tau}-\ref{eq:ro}): 
\begin{eqnarray}\label{eq:tau-i} 
\left(\partial_t - k_{(i)} \nabla^2 + \vec{v}_{(i)}\cdot \nabla \right) \tau_{(i)} &=& c^{-1}_{(i)} E_{(i)}(\tau_{(i\pm1)},\vec{p}_{(i\pm1)},\rho_{(i\pm1)}; t,\vec{x}) \;, \\
\label{eq:u-i}
\left(\partial_t +\eta_{(i)} \; +\; 2\vec{\omega} \times\; +\; \vec{v}_{(i)}\cdot \nabla \;\right) \vec{p}_{(i)} &=& \vec{F}_{(i)}(\tau_{(i\pm1)},\vec{p}_{(i\pm1)},\rho_{(i\pm1)})\;, \\
\label{eq:ro-i}
\left(\partial_t - \sigma_{(i)} \nabla^2 \right)\rho_{(i)} &=& \varphi_{(i)}(\tau_{(i\pm1)},\vec{p}_{(i\pm1)},\rho_{(i\pm1)}) \;,
\end{eqnarray}
where the label $i$ refers to a specific layer. We leave the study of the composite system for a later project.

\subsection{Perturbation series} 
Consider a solution of the form $\{\tau^{(0)}+\delta \tau, \vec{p}^{(0)} + \delta\vec{p}, \rho^{(0)}+\delta \rho\}$ where $\{\tau^{(0)}, \vec{p}^{(0)}, \rho^{(0)}\}$ is the solution of the linearized system, such as the one worked out in section 4 (zeroth order). This linearized solution determines the zeroth order velocity field $\vec{v}^{(0)}=\vec{p}^{(0)}/\rho_0$, where $\rho_0$ is the spacetime average of the unperturbed matter density. Then, the first order corrections $\{\delta \tau,\delta\vec{p},\delta \rho\}$ produced by the advective perturbation can be computed as 
\begin{eqnarray} \label{eq:deltatau}
(\partial_t -k\nabla^2 +\vec{v}^{(0)}\cdot\nabla)\delta \tau &=& - \vec{v}^{(0)}\cdot \nabla \tau^{(0)} \;,\\ \label{eq:deltau}
\left(\partial_t +\eta \; +\; \nu\,\hat{r} \times +\vec{v}^{(0)}\cdot\nabla \right) \delta \vec{p} &=&  - \vec{v}^{(0)}\cdot \nabla \vec{p}^{(0)} \; -R \nabla \delta \tau\;, \\
\label{eq:deltaro}
\left(\partial_t - \sigma \nabla^2  \right) \delta \rho  &=& \beta(\theta)\,\hat{r}\cdot (\nabla \times \delta \vec{p}) - \nabla \cdot \delta \vec{p} \;. 
\end{eqnarray}
The first order corrected solution can be plugged again in the system to compute the second order correction in an iterative process. The usefulness of the perturbative expansion is limited by the extent to which the emergent chaotic phenomena can be kept under control, but it is well known that this is in general an open problem.

Whether the advection terms can be safely neglected or one intends to carry out the perturbative analysis, it is necessary to solve the linear problem in either case.

\section{Summary and outlook}     
The approximately two-dimensional nature of the Earth's atmosphere at a global scale suggests the pertinence of its dynamical description as a fluid in two spatial dimensions. The main usefulness of the CS approach is that it selects the relevant field variables encoding the essential degrees of freedom of the system as well as the manner in which the dynamical variables relate to each other. In this case, those variables are the matter density ($\rho$), the momentum density ($\vec{p}=\rho \vec{v}$) and the thermal density which, for an ideal gas, is proportional to the pressure ($\tau= \rho T \propto P$). 
Including the external energy source, dissipation and diffusion gives a system of nonlinear coupled partial differential equations of first order in time (\ref{eq:tau}-\ref{eq:ro}). These equations account for the atmospheric fluid as an open dissipative system.\\

\noindent
A simplified linear version of this model is obtained assuming a single atmospheric layer and neglecting the advective derivatives, from which the resulting dynamics is described by an integrable set of linear differential equations. Thus, the integration can be completely expressed in analytic form for all values of the free parameters of the model.\\

\noindent
The energy function on the right hand side of \eqref{eq:tau} is the main input of the model. The parameters $k, c, R, \eta, \sigma, \beta, \lambda$ can be adjusted to describe different features of the atmosphere or even the atmospheres of different planets.
When focusing on the steady state, assuming the equilibrium condition in which the energy reaching the Earth equals that emitted to outer space, sets the global average temperature as  
\begin{equation}
T_0= \frac{E_0}{4\lambda\rho_0}\,.
\end{equation}
This formula allows to estimate the global average equilibrium temperature if any of these parameters change.\\

\noindent
The linearized single layer model on a uniform surface is admittedly a crude approximation that could be greatly improved to produce a more accurate picture. These improvements can include: considering two or three interacting layers instead of a single one, allowing for a non-uniform surface, etc. The resulting system will be necessarily more complex but still linear and accessible with similar methods to the one discussed here.
The inclusion of the advective derivative terms, on the other hand, would bring in nonlinearities (including the horizontal heat and momentum eddy fluxes) that could be treated perturbatively or numerically. This would improve the accuracy of the short-term description and could also result in long-term unpredictability (chaos) and possible instabilities. 
The accuracy of the numerical modelling would crucially depend on a correct adjustment of the free parameters of the model to their expected/observed values. A general question prompted by the full nonlinear problem is about the stability and the long-term predictability of the model. In the nonlinear regime, extreme sensitivity to the initial conditions leading to instabilities and chaos can be expected, reducing the usefulness of the model as a long-term predictor.\\

\noindent
Clearly the simulations presented here can be made more realistic in several ways:
\begin{enumerate}
\item Inclusion of geographic features. 
\item Considering a multilayer scenario.
\item Allowing $\Delta$ to oscillate with a period of one year. 
\item Allowing changes in $\lambda$ due to changes in the physical features of the atmosphere.
\end{enumerate}

\noindent
The inclusion of local effects would introduce additional small scale features coming from higher harmonics (larger $l$'s and $m$'s), generating more complex patterns. The multilayer approach could be more realistic but it could also lead to instabilities and chaos. The inclusion of adiabatic changes of $\Delta$ or $\lambda$ could be modelled by a parametric evolution as in a quasi-static form. For instance, varying $\lambda$ would result in a new equilibrium mean temperature of the atmosphere. A more extensive analysis of those simulations will be deferred for a future study.\\

\section*{Acknowledgments} 
We would like to thank Fabrizio Canfora, Nathalie Deruelle, Nicol\'as Donoso, Mikhail Kurgansky, Maisa Rojas, Olivia Romppainen-Martius, Roberto Rondanelli  and Andr\'es Sep\'ulveda for many enlightening comments and discussions. Our special thanks to Miguel Bustamante and F\'abio Novaes for their interesting critical comments, discussions and suggestions, to Cristi\'an Mart\'{\i}nez for his helpful advice with Mathematica, and to Ra\'ul Barriga for invaluable help with the technical issues and much more. This work has been partially supported by ANID/Fondecyt grants 11170486, 1180368 and 1220862, ANID/FB210021; and ANID/FONDAP/15110009; and by USS grant VRID-Inter22/10.
%

\begin{appendices}
\renewcommand{\theequation}{\thesection.\arabic{equation}}
\setcounter{equation}{0}
\section{Abelian CS equations in 3D}
\label{app:CS}

The three dimensional Chern-Simons action for the field $A$ (Abelian connection) in the presence of an external source $J$ can be written as \cite{Arnold,CS}
\begin{equation}
    I[A,j]= \int \left(\frac{1}{2} A\wedge dA -  J\wedge A \right)\;.
\end{equation}
Here $A$ is a one-form, $A_\mu dx^\mu = A_0 dt + A_i dx^i$ and $J$ is a two-form, $J_{\mu \nu} dx^\nu \wedge dx^\lambda$. The field equations are obtained by varying $I$ with respect to the field $A$, which yields 
\begin{equation}
F = J\; \qquad \mbox{or} \qquad  F_{\mu \nu} =J_{\mu \nu} \;,
\end{equation}
where $F=dA$ and, following the electromagnetic tradition, we define $F_{\mu \nu} := \partial_\mu A_\nu - \partial_\nu A_\mu$. A more familiar representation of $J$ in three dimensions is $J_{\mu \nu} = \frac{1}{2}\epsilon_{\mu \nu \lambda}j^\mu$, where $j$ is a three-component vector current density.

Consider a 2-dimensional sphere $S^2$ of radius $r_0$. Let $A=A_\mu dx^\mu$ be a 1-form in the spacetime $\mathcal{M} = \mathbb{R}\times S^2$, where $x^0 = t\in \mathbb{R}$ is time and $(x^1, x^2)$ are coordinates on $S^2$. The kinetic term of the CS Lagrangian is $ \epsilon^{\mu \nu \lambda} A_\mu \partial_\nu A_\lambda$, where $\epsilon^{\mu \nu \lambda}$ is the completely antisymmetric invariant Levi-Civita tensor, defined so that $\epsilon^{012}=+1$.

Interactions with external sources are described by the current density $_{\mu \nu}$ or its dual, $j^\mu \equiv \epsilon^{\mu \nu \lambda} J_{\nu \lambda}$. The field equations read
\begin{equation}
\partial_\mu A_\nu - \partial_\nu A_\mu = J_{\mu \nu}  = \epsilon_{\mu\nu \lambda} j^\lambda. \label{eq:interact-field}
\end{equation}
Separating the space and time components, the above expressions are
\begin{align}
\partial_0 A_i - \partial_i A_0 = J_{0 i}= \epsilon_{ik}j^k\; , \\
\partial_i A_j - \partial_j A_i = J_{ij} = \epsilon_{ij}j^0\; .
\end{align}

\section{Units and dimensions}  
\label{app:units}
The physical units of the various quantities involved here are $[\rho] = ML^{-2}$, $[v^i] = LT^{-1}$ and $[P]=M T^{-2}$, where $M$ is mass, $L$ is length, $T$ is time and $\Theta$ is temperature. The momentum density current $\vec{u}=\rho \vec{v}$ represents the mass crossing a unit area per unit time and its dimensions are  $[u]=ML^{-1}T^{-1}$. Therefore, the components of $A_\mu$ have the following dimensions (units), 
\[
[A_0] =[P]= M T^{-2} \; , [A_i] =[\rho v^i]= ML^{-1}T^{-1} \; .
\]
Thus, the dimensions of $A_0 dt$ and $A_i dx^i$ are the same ($M T^{-1}$), and the 1-form $A=A_0 dt + A_i dx^i$ is well defined. 

In order to match the units in equation \eqref{eq:interact-field}, the  dimensions of $j$ are $[j^0]=ML^{-2}T^{-1}$ and $[j^i]=ML^{-1}T^{-2}$. Then, 
\[
[AdA] =  [A_0j^0 d^3x] =  [ A_i j^i d^3x] = M^2 T^{-2}\;\; .
\]
The physical dimensions of the parameters that enter in the dynamical equations (\eqref{eq:tau}-\eqref{eq:ro}) are the following: 
$[\mathcal{E}_0] = M T^{-3}$, $[c] = L^{2} T^{-2} \Theta^{-1}$, $[r_0] = L$, $[R] = L^2 T^{-2} \Theta^{-1}$, $[\eta] = [\omega] = T^{-1}$, $[k] = [\sigma] = L^2 T^{-1}$, $[\lambda] =L^2 T^{-3} \Theta^{-1} $. In MKS units (m, kg, s, K), the basic parameters of the Earth are the following:

Earth radius: $r_0 = 6.4\times 10^6$ m.

Rotation frequency: $\omega=2\pi/$d $ \approx 7 \times 10^{-5}\, $s$^{-1}$.

Air density (at sea level): $\rho_{atm} =1.3\,$kg$/$m$^3$.

Average Earth's atmospheric temperature (tropopause) \cite{StandardAtmosphere}: $T_0 \approx 255\,$K.

Specific heat of air at constant pressure (at sea level, $ 300$K): $c_p=1004 $m$^2 $s$^{-2}$ K$^{-1}$.

Gas constant for dry air: $R=287 $m$^2 $s$^{-2} $K$^{-1}$.\\

In this model, a set of ``natural units" could correspond to choosing the Earth's radius $r_0$, the rotation frequency $\omega$ and the mean air density $\rho_h$ equal to 1. The remaining parameters of the system can be adjusted to model different scenarios as shown in Table 1. 
\begin{table}[h!]
\renewcommand{\arraystretch}{1.4}
\centering
\begin{tabular}{|c|l|c|c|}
\hline
\bf{Parameter}  &   {\bf Meaning}               & {\bf Default value}   & {\bf Range}\\ \hline
$r_0 $           &   Earth's radius              &   1   &   -   \\ \hline
$\omega$        &   Rotation frequency          &   1   &   -   \\ \hline
$\tau_0 $       &   Mean thermal density        &   1  &    -   \\ \hline
$\rho_s$        &   Mean two-dimensional density                &   1  &    -   \\ \hline
$\mathcal{E}_0$ &   Solar flux density          &   0.016    &   0 - 0.1  \\ \hline
$c$             &   Specific heat               &   1.14 &   0.1 - 2.0   \\ \hline
$\lambda$       &   Emissivity                  &   0.004 & 0 - 0.1    \\ \hline
$k$             &   Heat diffusion coefficient  &   0.03 &   0 - 1.0   \\ \hline
$R$        &   Ideal gas constant          &   0.33  &  0 - 0.5  \\ \hline
$\eta$          &   Viscous damping             &   0.5  &  0 - 1.0  \\ \hline
$\sigma$        &   Matter diffusion coefficient&   0.1  &  0 - 1.0  \\ \hline
$\beta_0$       &   Curl coefficient            &   $-$2.0  &  0 - $-$10  \\ \hline
\end{tabular}
\caption{Parameters of the model. The values of $r_0$, $\omega$, $\tau_0$ and $\rho_s$ have been set equal to 1, as they define the natural units for length, time, temperature and mass. The default values for $\mathcal{E}_0$, $c$ and $R$ were taken from the literature \cite{Wallace,Holton} and expressed in natural units. The values for $k$ and $\sigma$ are chosen so that the temperature in the polar regions does not blow up ($k > 0.01$ satisfies the P\'eclet condition). In the case of the damping coefficient $\eta$, it was set to represent an underdamped driven oscillator with little dissipation. Finally, the absolute value for $\beta_0$ is set to be near 1, which means giving similar weights to the curl and divergence contributions. The ranges are those implemented in \cite{App}.}
\label{tab:params}
\end{table}
\newpage
\section{Integration} 
\label{app:integration}
In the untilted case ($\Delta=0 \Rightarrow \hat{r}\cdot \hat{s} =\cos{\theta} \sin(\phi'+\omega t')$), the integral \eqref{eq:tau-lambda}, with the Green function $G_k$ in \eqref{eq:Gd} and the energy function is given by \eqref{eq:Ein}, 
\begin{align} \label{eq:tau-D} \nonumber
\tau(t,\theta,\phi) = \frac{\mathcal{E}_0}{c}\sum_{l,m} N_{lm} Y_{lm}(\theta, \phi) \int_{-\pi/2}^{\pi/2} d\theta'\cos^2{\theta'} P_l^m(\theta')  \int_0^t dt'\, e^{-Q_l(t-t')} \,\\
\times \int_0^{2\pi} d\phi'e^{-im \phi'} \sin(\phi'+\omega t')\, \Theta[\sin(\phi'+\omega t')]\, + \tau_h \;,
\end{align}
with $Q_l>0$, given in (\ref{44}). Since the last integral in $\phi'$ is invariant under $\phi'\to \phi'+a$, and the support of the Heaviside function is the set $0< \phi'+ \omega t'<\pi$, this last integral is
\begin{equation}
 \int_{-\omega t'}^{\pi-\omega t'} d\phi'e^{-im \phi'} \sin(\phi'+\omega t')= s_m \;e^{i m \omega t'},\; \mbox{with}\;\, s_m = \left\{
\begin{array}{ccl} 
\pm \frac{\pi}{2i} & \mbox{for} & m=\pm 1 \\
\frac{-2}{m^2-1} & \mbox{for} & \mbox{even } m \\
 0 & \mbox{for} & \mbox{odd } m \neq \pm 1
\end{array} 
\right. 
\end{equation}
Integrating over $t'$ and $\theta'$ yields
\begin{equation} \label{eq:tau''}
\tau(t,\theta,\phi) = \frac{\mathcal{E}_0}{c}\sum_{l,m} s_m\, \frac{\left[e^{i m \omega t} - e^{-Q_l t}\right]}{Q_l +i m \omega}
N_{lm} Y_{lm}(\theta, \phi)\,K_{lm} \; + \tau_h  \;,
\end{equation}
where $K_{lm}=\int_{-\pi/2}^{\pi/2} d\theta' \cos^2{\theta'} \,P_l^m(\theta')$.
Here the first term in brackets gives the steady state oscillatory contribution and the second term is the transient decaying mode. Finally, the steady state is
\begin{align}
\tau(t,\theta,\phi) = \frac{\mathcal{E}_0}{c}\sum_{l,m} \frac{s_m\, e^{i[ m (\phi+\omega t)-\delta_{lm}]}}{\sqrt{Q_l^2 + m^2 \omega^2}}\,K_{lm}\, N_{lm}^2 \,P_l^m(\sin{\theta}) \;,
\end{align}
where $\sin{\delta_{lm}}= m\omega/\sqrt{Q_l^2 + m^2 \omega^2}$. 

The derivatives of $\tau$ provide the sources for the components of the momentum densities according to (\ref{vecF1},\ref{vecF2}). 
The steady state solution for $\vec{p}$ is also a double summation where all the terms for odd $l$ and $m$, except $(l=1,m=\pm1)$, are zero.  Keeping the first of these harmonics ($l=0,1,2$), given by \eqref{eq:u_sol}, the steady state solution for $\vec{p}$ can be written as
\setlength{\jot}{12pt}
\begin{equation}
\SMALL{ 
\begin{split}
p^{\theta}(t,\vec{x}) = &\frac{15 \,R \,\mathcal{E}_0 \,\eta\,\sin {\theta}\cos{\theta}}{32 \,c\,r_0 \,Q_2 [\eta ^2 + (2 \omega \sin\theta)^2]} + \frac{R\, \mathcal{E}_0\,\sin{\theta} \sqrt{9\omega^2+\eta^2}\;\sin(\phi+\omega t +\epsilon_1(\theta))}{2 c r_0 \sqrt{(Q_1^2+\omega^2) (16 \omega^4 \sin ^4{\theta}-8 \omega^2 \sin ^2{\theta} (\omega^2-\eta ^2)+(\eta ^2+\omega^2)^2)} }\\ 
& + \frac{15 R \mathcal{E}_0 \sqrt{\eta ^2+16 \omega^2}\sin (2 \theta)\, \sin(2(\phi + \omega t)+\epsilon_2{\theta})}{64 c r_0 \sqrt{(Q_2^2+4 \omega^2) (16 \omega^4 \sin ^4{\theta} + 8 \omega^2 \sin ^2{\theta}(\eta ^2-4 \omega^2)+(\eta ^2+4 \omega^2)^2)}}\,,
\end{split}}
\end{equation}  
\begin{equation}
\SMALL{ 
\begin{split}
p^{\phi}(t,\vec{x}) &= \frac{15 R  \mathcal{E}_0 \omega \sin ^2{\theta} \cos{\theta}}{16 c Q_2 r_0 (\eta ^2+4 \omega^2 \sin ^2{\theta})} + 
\frac{R  \mathcal{E}_0 \sqrt{\eta ^2+\omega^2 (2-\cos (2 \theta ))^2} \sin (\phi + \omega t + \kappa_1(\theta))}{2 c r_0 \sqrt{Q_1^2+\omega^2} \sqrt{16 \omega^4 \sin ^4{\theta} - 8 \omega^2 \sin^2{\theta} (\omega^2-\eta ^2)+(\eta ^2+\omega^2)^2}} \\
  & + \frac{15 R  \mathcal{E}_0 \cos{\theta} \sqrt{\eta ^2+\omega^2 (3-\cos (2 \theta ))^2} \sin
   (2 (\phi+\omega t)+\kappa_2(\theta))}{32 c r_0 \sqrt{Q_2^2+4 \omega^2} \sqrt{16 \omega^4 \sin ^4{\theta}+8 \omega^2 \sin ^2{\theta} (\eta ^2-4 \omega^2)+(\eta ^2+4
   \omega^2)^2}} \,,
\end{split}}
\end{equation}
where  $\epsilon_l$ and $\kappa_l$ are the $\theta$-dependent phase shifts. 
Finally, plugging the above expression in \eqref{eq:rho_sol} yields the expression for $\rho(t,\vec{x})$,
\begin{equation}
    \rho(t,\theta, \phi) = \rho_0(\theta) + \sum_{\bar{l},l,m} \rho_{\bar{l},l m}\,,
\end{equation}
where $\rho_0$ is the time average of atmospheric density calculated as:
\begin{equation}
\rho_0(\theta) = \frac{r_0^2}{\sigma}\sum_{\bar{l}} N_{\bar{l}}^2 P_{\bar{l}}(\sin\theta) \int_{\pi/2}^{\pi/2} \cos{\theta'}P_{\bar{l}}(\theta')F_0(\theta')\theta'd\theta'
\end{equation}

The first harmonics ($l \leq 4$) of the inhomogeneous component (summation term) are
\begin{equation}
\SMALL{
    \begin{split}
\rho_0 &=  \mathcal{E}_0 R r_0^2 Z_2  P_0^0(\sin{\theta})\\
\rho_1 &=  \mathcal{E}_0 R r_0^2 \left[
0 + 
 \frac{P_1^1(\sin{\theta})(X_1(\theta)\cos(\phi+\omega t) + W_1(\theta)\sin(\phi+\omega t))}{c^2 r_0^8\,(Q_1^2 + w^2) (q_1^2 + w^2)}
 \right]\\
\rho_2 &=  \mathcal{E}_0 R r_0^2 \left[
 Z_2 P_2^0(\sin{\theta}) + 
\frac{P_2^1(\sin{\theta})(X_2(\theta)\cos(\phi+\omega t) + W_2(\theta)\sin(\phi+\omega t))}{c^2 r_0^8 \,(Q_1^2 + w^2) (q_2^2 + w^2)} \right. \\
& \qquad \qquad  \qquad  \qquad  \qquad + \left.
 \frac{P_2^2(\sin{\theta})(\bar{X}_2(\theta)\cos[2(\phi+\omega t)] + \bar{W}_2(\theta)\sin[2(\phi+\omega t)]}{c^2 r_0^8\,(Q_2^2 +(2 w)^2) \left(\frac{q_2^2}{4} + w^2\right)}\right]\\
 \rho_3 &=  \mathcal{E}_0 R r_0^2 \left[
0 + 
\frac{P_3^1(\sin{\theta})(X_3(\theta)\cos(\phi+\omega t) + W_3(\theta)\sin(\phi+\omega t))}{c^2 r_0^8\,(Q_1^2 + w^2) (q_3^2 + w^2)} \right. \\
& \qquad \qquad  \qquad  \qquad  \qquad + \left.
 \frac{P_3^2(\sin{\theta})(\bar{X}_3(\theta)\cos[2(\phi+\omega t)] + \bar{W}_3(\theta)\sin[2(\phi+\omega t)]}{c^2 r_0^8\,(Q_2^2 +(2 w)^2) \left(\frac{q_3^2}{4} + w^2\right)}\right]\\
 \rho_4 &=  \mathcal{E}_0 R r_0^2 \left[
 Z_4 P_4^0(\sin{\theta}) + 
\frac{P_4^1(\sin{\theta})(X_4(\theta)\cos(\phi+\omega t) + W_4(\theta)\sin(\phi+\omega t))}{c^2 r_0^8\,(Q_1^2 + w^2) (q_4^2 + w^2)} \right. \\
& \qquad \qquad  \qquad  \qquad  \qquad + \left.
 \frac{P_4^2(\sin{\theta})(\bar{X}_4(\theta)\cos[2(\phi+\omega t)] + \bar{W}_4(\theta)\sin[2(\phi+\omega t)]}{c^2 r_0^8\, (Q_2^2 +(2 w)^2) \left(\frac{q_4^2}{4} + w^2\right)}\right],
\end{split}}
\end{equation}
where $q_l=l(l+1)\sigma/r_0^2$, and the expressions for $X_l(\theta), W_l(\theta), Z_l(\theta)$,  $\epsilon_l(\theta)$ and $\kappa_l(\theta)$ can be requested to 
the corresponding author.\\
\end{appendices}

\end{document}